\documentclass{emulateapj}
\usepackage[utf8]{inputenc}
\usepackage{graphicx}
\usepackage{color}
\usepackage[dvipsnames]{xcolor}
\usepackage{amsmath}
\usepackage{esint}
\usepackage{soul}
\usepackage[normalem]{ulem}

\newcommand{\Msun}{\ensuremath{\,M_\odot}}
\newcommand{\Rsun}{\ensuremath{\,R_\odot}}
\newcommand{\Zsun}{\ensuremath{\,Z_\odot}}
\newcommand{\Lsun}{\ensuremath{\,L_\odot}}

\newcommand{\yr}{\ensuremath{\,\mathrm{yr}}}
\newcommand{\kyr}{\ensuremath{\,\mathrm{kyr}}}
\newcommand{\myr}{\ensuremath{\,\mathrm{Myr}}}
\newcommand{\gyr}{\ensuremath{\,\mathrm{Gyr}}}

\newcommand{\kms}{\ensuremath{\,\mathrm{km}\,\mathrm{s}^{-1}}}

\newcommand{\msy}{\ensuremath{\Msun\mathrm{\; yr}^{-1}}}

\newcommand{\pc}{\ensuremath{\,\mathrm{pc}}}
\newcommand{\kpc}{\ensuremath{\,\mathrm{kpc}}}

\newcommand{\microarcsec}{\ensuremath{\,\mu \mathrm{as}}}

\newcommand{\Ma}{\ensuremath{M_{\rm a}}}
\newcommand{\Mb}{\ensuremath{M_{\rm b}}}
\newcommand{\Mbh}{\ensuremath{M_{\rm BH}}}
\newcommand{\Mcomp}{\ensuremath{M_{\rm comp}}}
\newcommand{\Mzams}{\ensuremath{M_{\rm ZAMS}}}
\newcommand{\Mzamsa}{\ensuremath{M_{\rm ZAMS,a}}}
\newcommand{\Mzamsb}{\ensuremath{M_{\rm ZAMS,b}}}

\newcommand{\startrack}{{\tt StarTrack}}

\newcommand{\sci}[2]{\ensuremath{#1\times10^{#2}}}
\newcommand{\few}[1]{\ensuremath{\mathrm{a\,few}\times #1}}

\newcommand{\new}[1]{{\bf #1}}

\renewcommand{\new}[1]{#1}

\newcommand{\gaia}{{Gaia}}
\newcommand{\lamost}{{LAMOST}}

\newcommand{\std}{STD}

\newcommand{\ssa}{SS0}
\newcommand{\nkr}{NK$_{\rm R}$}
\newcommand{\nkbe}{NK$_{\rm BE}$}
\newcommand{\imff}{flatIMF}
\newcommand{\imfs}{steepIMF}

\defcitealias{Wiktorowicz1911}{W19}
\defcitealias{Olejak1908}{O19}

\begin{document}

\title{Non-interacting Black Hole binaries with Gaia and LAMOST}

\author{Grzegorz Wiktorowicz\altaffilmark{1,2}\thanks{E-mail: gwiktoro@astrouw.edu.pl},
        Youjun Lu\altaffilmark{1,2},
        \L{}ukasz Wyrzykowski\altaffilmark{3},
        Haotong Zhang\altaffilmark{1},
        Jifeng Liu\altaffilmark{1,2,4},
        Stephen Justham\altaffilmark{1,2,5},
        Krzysztof Belczynski\altaffilmark{6}}

 \affil{  
     $^{1}$ National Astronomical Observatories, Chinese Academy of Sciences, Beijing 100101, China\\
     $^{2}$ School of Astronomy and Space Science, University of the Chinese Academy of Sciences, Beijing 100012, China\\
     $^{3}$ Astronomical Observatory, Warsaw University, Al. Ujazdowskie 4, 00-478 Warsaw, Poland\\
    $^{4}$ WHU-NAOC Joint Center for Astronomy, Wuhan University, Wuhan, China\\
    $^{5}$ Astronomical Institute Anton Pannekoek, University of Amsterdam, P.O. Box 94249, 1090 GE, Amsterdam, The Netherlands\\
    $^{6}$ Nicolaus Copernicus Astronomical Center, Polish Academy of Sciences, Bartycka 18, 00-716 Warsaw, Poland \\
}

\begin{abstract}
    Until recently, black holes (BHs) could be discovered  only through accretion from other stars in X-ray binaries, or in merging double compact objects. Improvements in astrometric and spectroscopic measurements have made it possible to detect BHs also in noninteracting BH binaries (nBHBs) through a precise analysis of the companion's motion. In this study, using an updated version of the \startrack\ binary-star population modeling code and a detailed model of the Milky Way (MW) galaxy we calculate the expected number of detections for \gaia\ and \lamost\ surveys. We develop a formalism to convolve the binary population synthesis output with a realistic stellar density distribution, star-formation history (SFH), and chemical evolution for the MW, which produces a probability distribution function of the predicted compact-binary population over the MW.  This avoids the additional statistical uncertainty that is introduced by methods that Monte Carlo sample from binary population synthesis output to produce one potential specific realization of the MW compact-binary distribution, and our method is also comparatively fast to such Monte Carlo realizations. Specifically, we predict $\sim41$--$340$ nBHBs to be observed by \gaia, although the numbers may drop to $\sim10$--$70$ if the recent ($\lesssim100\myr$) star formation is low ($\sim1\msy$). For \lamost\ we predict $\lesssim14$ detectable nBHBs, which is lower partially because its field of view covers just $\sim6\%$ of the Galaxy.
\end{abstract}


\section{Introduction}

Black holes (BHs), by definition, are very hard to detect electromagnetically \footnote{Theoretically, they produce Hawking radiation \citep{Hawking7403,Hawking7508}}. Recently, the Event Horizon Telescope collaboration has detected a silhouette of a supermassive BH \citep{EHT1904} and merging BHs have been detected through gravitational wave emission \citep{Abbott1602}. Some methods like microlensing \citep[e.g.][]{Minniti1509,Wyrzykowski1605,WyrzykowskiMandel2019, Wiktorowicz1911,Masuda1910}, tidal disruption events \citep[e.g.][]{Perets1606,Kremer1908}, or accretion from dense interstellar medium \citep[e.g.][]{Tsuna1806} provide a way of detecting free-floating BHs. BHs bound in binaries may be detected through interactions with their companion stars. If the separation is small enough that the companion is able to fill its Roche lobe (RL) during its evolution, stable mass transfer (MT) may occur and the system will become observable as an X-ray binary \citep[e.g.][]{Novikov7301,Wijnands9811,Zdziarski0401,Gilfanov0403,Zycki0505,Maccarone0511,Fabbiano0609,Middleton1203,King1410,Tetarenko1602}, although some systems can be hard to detect owing to low luminosities \citep[e.g.][]{Menou9903}. When the separation ($a$) is larger, the MT through RL overflow (RLOF) cannot occur, but the X-ray emission may be powered by wind accretion. However, the MT through stellar wind is typically small \citep[but see, e.g.][]{Mohamed1201,Liu1311,elMellah1902}, and the emission is in general weaker than from RLOF-fed systems. Nonetheless, even in the absence of interactions, the orbital motion of the visible star detected either astrometrically or spectroscopically may indicate the presence of an invisible companion \citep[e.g.][]{Casares1401,Giesers1803,Liu1911,Thompson1911}. If, additionally, the mass of the hidden object is estimated to be large ($M\gtrsim3\Msun$), then we have a strong claim for a BH, because a regular star of such a mass should be very luminous \citep[e.g.][]{Karpov0110,Yungelson0608}. Detection of these noninteracting BH binaries (nBHB) in ongoing and future surveys is a promising way of estimating the BH population of our Galaxy.

The \gaia\ mission of the European Space Agency \citep{Gaia1611}, with its unprecedented astrometric precision and number of observed stars, will provide a perfect database for a variety of statistical studies. 
Gaia scans the entire sky over the period of 5 \yr\ and delivers multiepoch photometric and astrometric observations of more than a billion stars. 
Regular subsequent data releases \citep{GaiaDR1, GaiaDR2} have provided, among many other products, positional parameters (positions, parallax, and proper motions) for nearly all monitored stars; however, the individual time series of the astrometric data will be released in the final data release, opening a new vault for scientific exploration. Gaia astrometry was already claimed to be useful in detecting invisible companions \citep[e.g.][]{Gould0206,Tomsick1008,Barstow1407,Igoshev1907}. Recently, \citet{Gandhi2009} used astrometric noise in the DR2 data release as a proxy for orbital motion in unresolved binaries identifying viable candidates for in-depth observations.

\citet{Kawanaka1701} and \citet{Mashian1709} made a proof-of-concept analytical estimation of the expected number of nBHB systems potentially detectable by \gaia\ through its $5$ \yr\ long mission. \citet{Yamaguchi1807} improved significantly their methodology considering, particularly, an interstellar absorption and obtained prediction of $200$--$1000$ nBHBs discoverable in the \gaia\ data. The method was further employed by \citet{Yalinewich1811}, who added the treatment of natal kicks \citep[NKs; e.g.][]{Herant9505,Hobbs0507,Fryer0604,Fryer0704,Kuznetsov1207}, the binary fraction, and the simple spatial distribution model, obtaining a prediction of $50$--$150$ nBHBs. Additionally, \citet{Shikauchi2006} found through $N$-body simulations that $\sim10$ nBHBs formed in open clusters may be present in the \gaia\ data.

All the previous research, however, suffers from a significant drawback, which is the lack of any treatment of binary interactions that may affect the predecessors of nBHBs \citep[hereafter \citetalias{Wiktorowicz1911}]{Wiktorowicz1911}.
The first study of these, which included binary interactions through employing the population synthesis (PS) method, was \citet{Breivik1711}, who derived an estimate of $3,800$--$12,000$ nBHBs, depending on the assumed \gaia\ astrometric precision. Additionally, their work included an improved Milky Way (MW) model and took into account that we actually observe projected orbits. Recently, \citet{Shao1909} performed a similar calculation using different code and model assumptions and estimated several hundred potential detections for \gaia.

Another way of detecting BH companions in nBHBs is the measurement of radial velocity (RV) variations through spectroscopic observations \citep[e.g.][]{Trimble6906,Giesers1803,Khokhlov1804,Makarov1904,Thompson1911}. One of the contemporary instruments devoted to spectroscopical observations is \lamost, which has $4000$ fibers and can take spectra of thousands of stars in a single observation. Therefore, it is perhaps the best instrument that can be used to search for spectroscopic binaries by monitoring numerous stars over a long period.

Recently, \citet{Gu1902} used \lamost\ DR6 and investigated six red giants (RGs) with detected high ($>80\kms$) RV variations. Their results show that on the basis of available data the presence of a BH primary cannot be rejected for all of these stars. The same type of systems was a target of the \citet{Zheng1911} study, which utilized \lamost\ data supported by ASAS-SN photometry, but also with no clear detections. \citet{Yi1912} calculated the predicted detection rate of nBHBs by \lamost. They used a method similar to \citet{Mashian1709} and have used simple substitutions to binary formation and stellar evolution process, meanwhile totally ignoring the binary interactions. Their result claimed $50$--$400$ nBHBs potentially observable by \lamost.

In this work we want to significantly improve previous predictions using publicly available\footnote{\it https://universeathome.pl/universe/bhdb.php} models of BH populations in different stellar environments \citepalias{Wiktorowicz1911}. We predict the number of nBHBs present in the M, with a particular attention paid to these that will be observable by \gaia\ and \lamost. In Section~\ref{sec:methods} we describe the utilized database, the model for the MW, and limitations that were imposed on the synthetic results in order to obtain observational predictions. Section~\ref{sec:results} is dedicated to the presentation of the results and comparison of different models. In Section~\ref{sec:discussion} we compare our results with previous studies and discuss the problem of interpreting an observation as an nBHB. The summary is provided in Section~\ref{sec:summary}.

\section{Methodology}\label{sec:methods}

\subsection{Binary evolution models}\label{sec:binary_evolution}

\begin{deluxetable}{cl}
    \tablewidth{\columnwidth}
    \tablecaption{Summary of models}

    \tablehead{ Model & Difference with Respect to Standard Model }
    \startdata
    \std & Standard (reference) model: \\
     & \hspace{0.5cm}Distribution of initial periods\\
     & \hspace{0.5cm}\hfill $P(\log P)\propto(\log P)^{-0.55}$\\
     & \hspace{0.5cm}Distribution of initial eccentricities\\
     & \hspace{0.5cm}\hfill $P(e)\propto e^{-0.42}$\\
     & \hspace{0.5cm}BH/NS NKs are drawn from \\
     & \hspace{0.7cm}Maxwellian distribution with $\sigma=265$ km s$^{-1}$\\
     & \hspace{0.7cm}BH NKs reduced by fallback\\
     & \hspace{0.5cm}Moderate slope for high-mass end of the IMF\\
     & \hspace{0.5cm}\hfill $\Gamma=-2.3$\\
    \ssa & Distribution of initial separations $P(\log a)\propto 1$\\
     & Distribution of initial eccentricities $P(e)\propto e$\\
    \nkr\ & BH NKs are inversely proportional to the BH's mass \\
    \nkbe & BH/NS NK proportional to \\
     & \hfill ratio of ejecta mass and remnant mass\\
    \imff & Flat slope for high-mass end of the IMF ($\Gamma=-1.9$)\\
    \imfs & Steep slope for high-mass end of the IMF ($\Gamma=-2.7$)\\
    \enddata
    \tablecomments{All main parameters are provided only for the \std\ model. For other models, only differences in respect to the \std\ model are given explicitly. Reproduced from \citet{Wiktorowicz1911}.}
\label{tab:models}
\end{deluxetable}

We use the publicly available database of BHs in different stellar environments \citepalias{Wiktorowicz1911}. Their calculations  were performed with the recent version of the \startrack\ PS code \citep{Belczynski0206,Belczynski0801,Belczynski2004}. We take 18 models from this database: six of their main models (see Table~\ref{tab:models}): \std, \ssa, \nkr, \nkbe, \imff, and \imfs, which were calculated for solar metallicity ($Z=\Zsun=0.02$), and the same models but with different values of metallicity ($Z=10\%\Zsun=0.002$, and $Z=1\%\Zsun=0.0002$)\footnote{Specifically, models STD, SS0, NKr, NKbe, flatIMF, steepIMF, lowZ, lowZ\_SS0, lowZ\_NKr, lowZ\_NKbe, lowZ\_flatIMF, lowZ\_steepIMF, midZ, midZ\_SS0, midZ\_NKr, midZ\_NKbe, midZ\_flatIMF, midZ\_steepIMF.}. This allowed us to account for the metallicity distribution in the Galaxy for each of the models. From now on, where we talk about a particular model, we talk about three metallicity models combined. Below, we summarize the most important properties of the models \citepalias[][see Tab.~1 and Section~2 for further details]{Wiktorowicz1911}.

\citetalias{Wiktorowicz1911} performed simulations of homogeneous populations of stars with the same initial metallicity and no imposed star formation history (SFH). Their results may be therefore used as "building blocks" for more complicated stellar systems like galaxies, what we do here for the MW. The binaries were evolved in isolation of what is a justified argument for such sparse stellar systems like galactic disks \citep[although see][]{Klencki1708}. We note that interactions between stars and binaries can be important in dense stellar systems like the galactic nuclei, or globular clusters \citep[e.g.][]{Morawski2018}. The interactions between stars in binaries, which were included in the simulations of \citetalias{Wiktorowicz1911}, may in general affect the evolution of an nBHB predecessor. Such interactions include tidal interactions, NKs imparted on the compact object after formation, MT phases, common envelope (CE), etc. All of them, especially NKs, may change the properties of the population significantly and cannot be neglected. 

In Table~\ref{tab:models} we summarize the models used in the study for the convenience of the reader \citepalias[the models are discussed thoroughly in][]{Wiktorowicz1911}. The standard model (\std) is the most conservative one. The initial binary parameter distributions of orbital periods and eccentricities follow these observed by \citet{Sana1207}, but we have extended them to include mass ratios lower than $0.1$ and the full range of primary masses ($0.08$--$150\Msun$). The initial primary masses were drawn from the broken-power-law distribution with power-law index $\Gamma=-1.3$, $-2.2$, and $-2.3$ for mass below $0.5\Msun$, between $0.5$ and $1.0\Msun$, and above $1\Msun$, respectively. Mass ratios were drawn from uniform distribution, securing the lower limit for the secondary mass at $0.08\Msun$. The NS and BH NKs are drawn from the Maxwellian distribution with $\sigma=265\kms$ \citep{Hobbs0507}. In the case of BHs, we reduce the kicks proportionally to the fallback (i.e. we multiply them by $1-f_{\rm fb}$, where $f_{\rm fb}$ is the fraction of mass that falls back onto the compact object).

Other models are variations of the \std\ model. In the SS0 model the distribution of initial separations is flat in logarithm ($P(\log a)\propto 1$ with $a_{\rm max}=10^5\Rsun$), whereas the eccentricities have initially thermal distribution ($P(e)\propto e$). Models \nkr\ and \nkbe\ introduce new models for NKs. The \nkr\ model gives kicks inversely proportional to BH mass $v_{\rm kick}\propto M_{\rm BH}^{-1}$ \citep{Rodriguez1611}, whereas the NK in the \nkbe model is proportional to the ratio of ejecta and remnant mass $v_{\rm kick}\propto M_{\rm ejecta}/M_{\rm BH}$ \citep{Bray1811}. The flatIMF and steepIMF models present variations in the steepness of the initial mass function (IMF) for massive stars ($M_{\rm ZAMS}>1\Msun$) with $\Gamma=-1.9$ and $-2.7$, respectively.

\subsection{Milky Way model}\label{sec:MWmodel}

\begin{deluxetable*}{cccccccl}
    \tablewidth{\textwidth}
    \tablecaption{Models of the Milky Way components}

    \tablehead{ Component & $M_{\rm tot}$ [\Msun] & SFR [$\frac{\Msun}{\mathrm{yr}}$] & $t_{\rm SF, start}$--$t_{\rm SF, end}$ ($\gyr$) & $Z$ & $\frac{dN}{dV}$ & }
    \startdata
        Thin disk & \sci{4.7}{10} & 4.7 & 10--0 & \Zsun & $\exp\left(-(a/5)^2\right)-\exp\left(-(a/3)^2\right)$ & if $t_{\rm age}\leq0.15\gyr$\\
          &  & & & & $\exp\left(-\sqrt{0.25+(a/2.53)^2}\right)+$ \hfill$\,$\\
          & & & & & \hfill$-\exp\left(-\sqrt{0.25+(a/1.32)^2}\right)$ & if $t_{\rm age}>0.15\gyr$\\
        Thick disk & \sci{5}{9} & 2.5 & 11--9 & \Zsun/10 & $\exp(-\frac{R-\Rsun}{2.5}\times\left(1-\frac{z^2}{0.8}\right)$ & if $|z|\leq0.4$\\
           & & & & & $\exp\left(-\frac{R-\Rsun}{2.5}\right)\times1.32\times\exp\left(-\frac{|z|}{0.8}\right)$ & if $|z|>0.4$\\
        Halo & \sci{1}{9} & 0.5 & 12--10 & \Zsun/100 & $\left(\frac{0.5}{\Rsun}\right)^{-2.44}$ & if $a\leq0.5$\\
         & & & & & $\left(\frac{a}{\Rsun}\right)^{-2.44}$ & if $a>0.5$\\
        Bulge & \sci{9.1}{9} & 2.3 / 0.45 & 12--10 / 10--0 & \Zsun & $\exp\left(-0.5\times r_{\rm s}^2\right)$ & if $R\leq2.54$\\
        & & & & & $\exp\left(-0.5\times r_{\rm s}^2\right)\times\exp\left(-2\left(R-2.54\right)^2\right)$ & if $R>2.54$\\
    \enddata
    \tablecomments{The table presents the model MW components used for the presented study. Presented are $M_{\rm tot}$--total stellar mass of the component; SFR--star formation rate; $t_{\rm SF, start/end}$--look-back time of the start and end of star formation episode, which is also the range of ages of the stars in the component; $Z$--metallicity ($\Zsun=0.02$ is the solar metallicity); $dN/dV$--normalized number density of stars \citep[based on][]{Robin0310}; $x$, $y$, $z$ - galactocentric position coordinates in \kpc\ (solar position was assumed to be at $x_\odot$, $y_\odot$, $z_\odot$=8.3, 0, 0); $\Rsun=8.3$ is the distance of the Sun from the Galactic center in \kpc;  $R=\sqrt{x^2+y^2}$; $a=\sqrt{R^2+z^2/\epsilon^2}$, where $\epsilon=0.0551$ for the thin disk and $\epsilon=0.76$ for the halo; $r_{\rm s}=\sqrt{\left[\left(x/1.59\kpc\right)^2+\left(y/0.424\kpc\right)^2\right]^2+\left(z/0.424\kpc\right)^4}$; SFR was assumed to be constant during the star formation phase. The values for SFR and $t_{\rm age}$ were taken from \citet{Olejak1908}. For the bulge, there are two star formation episodes}
    \label{tab:MWcomponents}
\end{deluxetable*}

The detailed structure of the MW and, especially, its past evolution is still not well constrained. The MW galaxy is believed to have a complex structure similar to a barred spiral galaxy. The main components consist of a thin disk and a thick disk, a bulge, and a halo. The observations proved that different components differ not only in the spatial distribution of stars but also in the SFH and chemical composition. Here we model each of the mostly homogeneous components separately in order to obtain a realistic model of the MW and its history.  The usefulness of describing complex stellar systems with small building blocks was already pointed out by \citet{Bahcall8112}

In Table \ref{tab:MWcomponents}, we present the model used for the spatial distribution of stars in the Galaxy. We use normalized stellar number density distributions of stars provided by \citet{Robin0310} as a proxy for the spatial distributions of nBHBs. Table \ref{tab:MWcomponents} also provides metallicities and SFHs (both star formation rates and intervals), which are slightly simplified versions of those in \citet[][hereafter \citetalias{Olejak1908}]{Olejak1908}. The modification of their metallicity distribution is motivated by the range of models available in \citetalias{Wiktorowicz1911}. \citetalias{Olejak1908} based their model on observational data \citep[e.g.,][]{Bullock0512,Soubiran0507} and cosmological simulations \citep[e.g.][]{Kobayashi1103} and it treats the chemical evolution in more detail \new{\citep[see also][]{Vos2003}}. We note that the SFH in the Galaxy is actually poorly known and may differ by an order of magnitude depending on the observational method used for the constraints \citep[e.g.][]{Chomiuk1112}.

The duration of the star formation in our model is exactly the same as in \citetalias{Olejak1908}. The star formation rate in the thin disk we choose to be constant \citep[but see \citeauthor{Romano1011} \citeyear{Romano1011}]{Mutch1108} and equal to $4.7\msy$ so that the total mass of the disk (thin and thick) is $\sci{5.2}{10}\Msun$ \citep[see $\sci{5.17}{10}\Msun$;][]{Licquia1506}. As far as the bulge is concerned, we slightly lowered the star formation rate during the constant star formation phase to $4.5\msy$, in order to get the total mass of the bulge equal to $\sci{9.1}{9}\Msun$ \citep[see][]{Licquia1506}. Also, we chose the star formation rate in the halo to be higher, so the total stellar mass is $\sci{1}{9}\Msun$ to account for recent measurements using the \gaia\ data \citep[e.g.][]{Belokurov1807,Helmi1810}.

The metallicity distribution in the disk is chosen to be constant during the cosmic time in contrast to \citetalias{Olejak1908} who provide a relation between the metallicity and the stellar age. We also change the metallicities in MW components to fit models available from \citetalias{Wiktorowicz1911}. Specifically, we choose the thin disk to have solar metallicity $Z=\Zsun$ regardless of the age of stars. The thick-disk metallicity is chosen to be $\Zsun/10$, which is $\sim2$ times lower than in \citetalias{Olejak1908}. For the bulge, we assume the solar metallicity both during the burst and during the following constant star formation. The metallicity in the halo we adopt as constant and equal $\Zsun/100$ through the duration of star formation, which is assumed to have occurred in a single star formation burst $\sim10\gyr$ ago, which is in agreement with recent estimations made using \gaia\ data \citep[e.g.][]{Helmi1810,Matteucci1908}. See \citetalias{Olejak1908} and references therein for more details. 

We choose to impose an outer limit on the Galaxy at the distance $R_{\rm lim}=30\kpc$ from its center. It is motivated by observations \citep[e.g.][]{Sesar1104} which show that at $R_{\rm lim}\approx30\kpc$ the density gradient becomes very steep. The halo may actually extend to hundreds of kiloparsecs, but the fraction of stellar mass outside of $R_{\rm lim}$ is negligible, and, actually, few stars are found outside of $R=15\kpc$. 

Although it was shown that the disk and the halo may exchange mass \citep[e.g.][]{Fox1910}, no MT between the Galactic components is included explicitly in our model. Undetected mass exchange may affect the SFH, but the presented analysis, which depends solely on the rate of star formation, not its sources, is unaffected by this process. Also, mass accretion from intergalactic medium \citep[e.g.][]{Oort6912} is neglected.

We assume $R_\odot=8.3\kpc$  \citep{Gillessen0902,deGrijs1611}. The thin-disk mass estimation is very sensitive to the chosen value of $R_\odot$ \citep{Bovy1312}, so we choose a value consistent with \citet{Robin0310}. For practical reasons, we assume that the Sun is located at the galactic plane ($z_\odot=0$). The Sun may actually reside slightly above the Galactic plane \citep[e.g. $z_\odot\approx20\pc$;][]{Yoshii13}, but such a small deviation will not influence our results noticeably. Also, we choose the location of the x-axis and y-axis in the employed Cartesian coordinate system in such a way that $x_\odot=8.3\kpc$ and $y_\odot=0$. 

\begin{figure}
    \centering
    \includegraphics[width=\columnwidth]{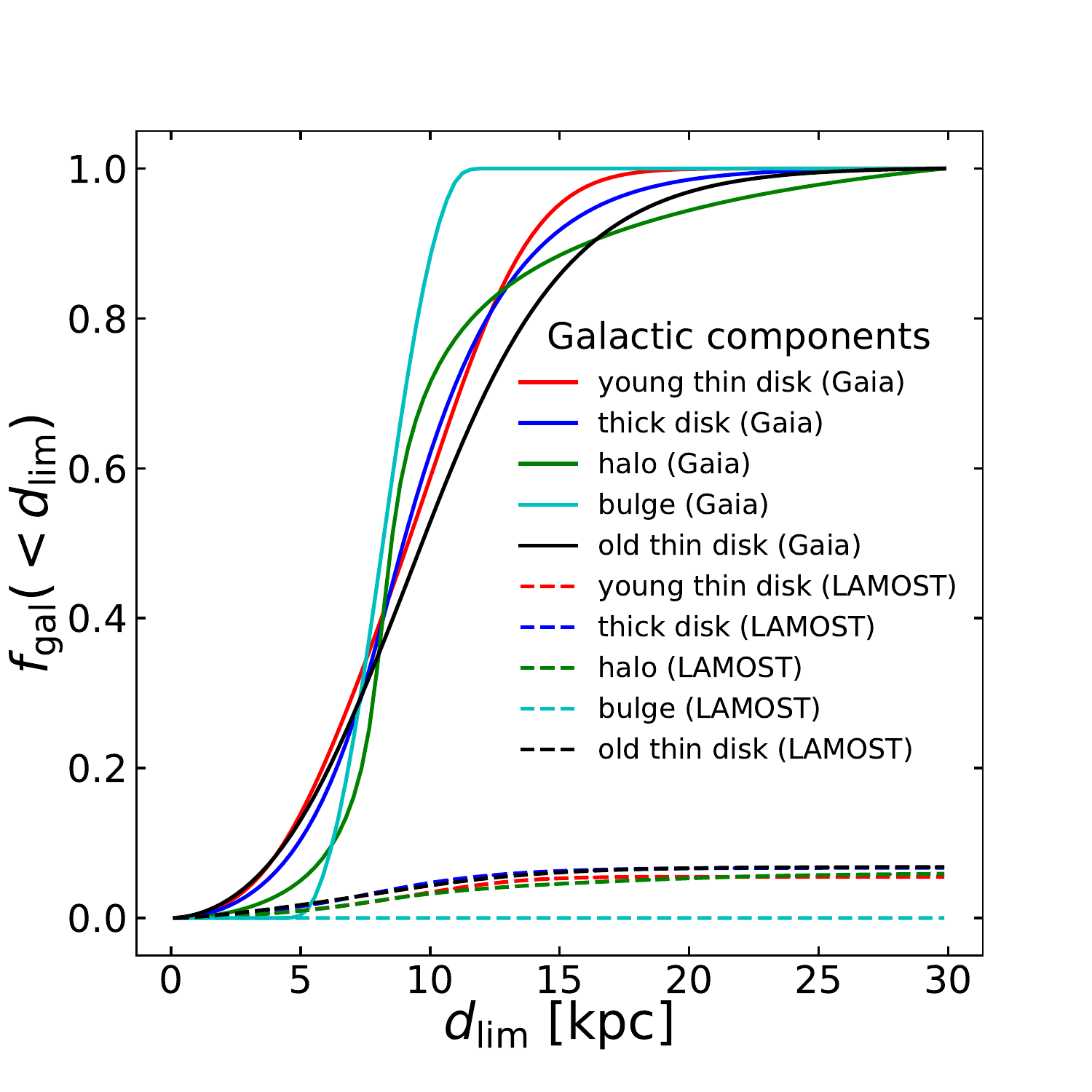}
    \caption{Fraction of the MW galaxy component stellar mass ($f_{\rm gal}$) within the distance $D$ from the Sun in the \gaia\ (solid lines) and \lamost\ (dashed lines). The \lamost\ field of view is limited to declinations $-10^\circ$ to $65^\circ$ and right ascensions $0^\circ$--$280^\circ$ which covers $\sim5$--$7\%$ of the Galaxy. Different lines represent different Galactic components (see Table~\ref{tab:MWcomponents}).}
    \label{fig:profiles}
\end{figure}

For the purpose of this study and to facilitate future research, we have calculated cumulative distributions of stellar mass as a function of distance from the Sun ($f_{\rm gal}$) with the division on different Galactic components (Figure~\ref{fig:profiles}). Precisely,
\begin{flalign}\label{eq:f_gal}
        \nonumber f_{\rm gal,\gaia}&(<D)=\\\nonumber&\iiint_{\sqrt{(x-x_\odot)^2+y^2+z^2}<D}\frac{dN}{dV}(x,y,z)dxdydz,\\
        f_{\rm gal,\lamost}&(<D)=\\\nonumber&\iiint_{\begin{subarray}{c}\sqrt{(x-x_\odot)^2+y^2+z^2}<D\\\delta\in[-10^\circ,65^\circ]\\\alpha\in[0^\circ, 280^\circ]\end{subarray}}\frac{dN}{dV}(x,y,z)dxdydz,
\end{flalign}    
\noindent where $dN/dV$ is the stellar number density defined in Table~\ref{tab:MWcomponents}, $D$ is the distance from the Sun, and the integration goes over a sphere centered at the Sun $\vec{r}_\odot=(x_\odot,y_\odot,z_\odot)=(8.3\kpc, 0, 0)$. We assumed that the edge of the Galaxy is located at $D_{\rm max}=30\kpc$ from the Galactic center and no stars are present outside of this radius \citep[see, e.g.,][]{Sesar1104}. As \lamost\ observes only a fraction of the sky, $f_{\rm gal}$ for \lamost\ is always lower than for \gaia\ $f_{\rm gal, \gaia}$, which observes the entire sky; \lamost's limited field of view (declinations $-10^\circ$ to $65^\circ$ and right ascensions $0^\circ$--$280^\circ$), which covers only $5$--$7\%$ of the Galactic stars depending on the Galactic component (except the bulge, where the fraction is $\sim0\%$; see Table.~\ref{tab:LAMOST_FOV}), must be included in the integration. The low $f_{\rm gal}$ for \lamost\ results mainly from not having the bulge and its surroundings (Galactic cener coordinates: $\delta_{\rm GC}=-29^\circ$ and $\alpha_{\rm GC}=195^\circ$) in its field of view, where the stellar density is the highest. We note that $f_{\rm gal}$ for \lamost\ is \emph{not} just a scaled-down $f_{\rm gal}$ for \gaia\ and, when normalized, has a generally different shape. It comes from the fact that the stellar density changes differently with distance depending on the direction of observation. The exemplary case is the direction toward the Galactic center and in the opposite direction.

\begin{deluxetable}{ccc}
    \tablewidth{\columnwidth}
    \tablecaption{\lamost\ Field of View}
    \tablehead{ Component & fraction & }
    \startdata
        Thin disk & $0.055$ & if $t_{\rm age}\leq0.15\gyr$\\
            & $0.068$ & if $t_{\rm age}>0.15\gyr$\\
        Thick disk & $0.067$ \\
        Halo & $0.059$ \\
        Bulge & $\lesssim10^{-6}$ \\
    \enddata
    \tablecomments{The star number fraction of each Galactic component observable by \lamost. $t_{\rm age}$ is the age of stars in the component. In contrast to \gaia, \lamost\ is able to observe only a part of the sky where decl. is $\delta\in[-10^\circ, 65^\circ]$ and right ascension $\alpha\in[0^\circ, 280^\circ]$, which results in $\sim5$--$7\%$ of each Galactic component being in the field of view, except the bulge, which is totally outside of \lamost's observational capabilities (a nonzero fraction results only from the lack of any strict size limit imposed on the Bblge while using the density formula [Table~\ref{tab:MWcomponents}], but has no effect on the results).
    }
    \label{tab:LAMOST_FOV}
\end{deluxetable}

\subsection{Observational cut}\label{sec:obscut}

In a widely utilized approach, the observational cuts are obtained by sampling the external parameter (i.e. not inherent to a binary)  distributions (e.g. spatial distributions, distribution of orbital orientations) for all the objects in a relevant population and checking whether the objects on which such parameters are imposed are observable, i.e., comply with all the limitations of a chosen instrument (the observable systems compose, so-called, observational cut). Such an approach can potentially lose a significant amount of information about the systems, since only some combinations (randomly chosen from distributions) of systems' extrinsic parameters are represented in the sample. In most cases these are the systems with the most typical parameters, which is what is wanted; however, some more rare configurations may occur by chance, which may potentially give a high weight to systems that in real situations are very rare, or in the opposite situation, rare (but interesting) systems may not appear in the results at all. To mitigate this problem, typically, many executions of Monte Carlo--based algorithms are necessary in order to obtain a satisfactory precision and get rid of ``artifacts.''

Here we present a novel approach to calculate the estimated number of observed sources from PS results that are free from the above problems. The basic idea is to change the stochastic sampling of extrinsic parameter distributions (esp. those relevant to the observational cut) by multidimensional integration of related probability distributions. The method avoids potential information leaks by using all the data concurrently and omits the introduction of additional statistical uncertainties (in contrast to Monte Carlo--based methods). Specifically, instead of drawing the extrinsic properties of systems and calculating the visibility indicators as Boolean variables, we propose to calculate the visibility as a probability by integrating normalized distributions of extrinsic parameters. For example, in the case of the spatial distribution, the probability corresponds to the fraction of the Galactic stellar mass within a range where the system is observable from the Earth. 

The probability corresponding to a particular system is, de facto, equivalent to the expected number of systems in the realistic population that are similar (have the same intrinsic properties as the synthetic system). The sum of expectations calculated for all systems in the synthetic sample obtained from PS simulations gives the expected number of considered systems in the observational results (after multiplication by the scaling factor $f_{\rm scale}$, see below). The main strength of this approach is that the procedure produces significantly smaller statistical uncertainties, which originate only from the numerical integration and not from the statistical noise. Consequently, it allows us to obtain stronger constraints on theoretical models. As most of the integration can be done in preprocessing by calculating cumulative distributions of extrinsic parameters, the additional computational cost of this method is constant, and for large data sets the method performs similarly or even better than Monte Carlo--based methods, especially if the former introduces several samplings of the synthetic data set. Below we present how this approach can be applied practically to provide estimations for \gaia\ and \lamost.

\subsubsection{Procedure for \gaia}\label{sec:nBHB_Gaia}

To ensure that all orbits are complete, we limit the periods to $P_{\rm orb,max}=5\yr$. We note that orbital parameters may be obtained even for incomplete orbits when the observed arc covers at least $75\%$ of the orbit \citep{Aitken18}. Recently, \citet{Lucy1403} reported that, using Bayesian analysis, it is possible to obtain orbital parameters when the orbit coverage is as low as $\sim40\%$. Such incomplete orbits may potentially increase the number of predicted detections, but they are not analyzed in this study. 

The estimated number of systems visible by \gaia\ can be obtained through the following formula:
\begin{flalign}\label{eq:nBHB_Gaia}
    \nonumber &N_{\rm nBHB, \gaia}=\\
    &\sum_{\rm nBHBs}\int_{\Delta t_{\rm nBHB}}dt\int_{b_{\rm comp}}^{a_{\rm comp}}da_{\rm proj}P(a_{\rm proj})f_{\rm w} P(d\leq d_{\rm max})),
\end{flalign}
\noindent where the summation goes through all the nBHBs obtained from the PS simulations \citepalias{Wiktorowicz1911}. Parameters of the binary ($a$,$e$,$\Mbh$, $\Mcomp$, $Z$, $L$, etc.) are obtained from the PS results and in general depend on evolutionary time $t$. $\Delta t_{\rm nBHB}$ is the time spent by the system as an nBHB. In the general case, the parameters of the system may change during an nBHB phase, and there can be more than one nBHB phase divided by interaction phases like MT or CE. The integral over the evolutionary time ($t$) includes these binary parameter changes. 

The observed orbit is the orbit of the companion whose semi-major axis ($a_{\rm comp}$) relates to the binary separation ($a$) as $a_{\rm comp}=a \Mbh / (\Mbh + \Mcomp)$, where $\Mbh$ and $\Mcomp$ are the mass of the BH and its companion, respectively. $a_{\rm proj}$ is the semi-major axis of the projected orbit on the plane perpendicular to the line of sight and including the original orbit's center of mass (we mark this plane as $p_{\rm obs}$). The plane of the original orbit ($p_{\rm orb}$) is in general inclined to $p_{\rm obs}$. To include the projection effect, we calculate the projected orbit on $p_{\rm obs}$ as \citep[see][]{Murray99}
\begin{equation}\label{eq:r_obs}
    \vec{r}_{\rm obs}=\vec{r}\left(\begin{array}{c}\cos(\omega + f)\\\sin(\omega + f)\cos(i)\end{array}\right),
\end{equation}
\noindent where $r=a(1-e^2)/(1+e\cos f)$ describes the original orbit for a parameter $f$, $\omega$ is the position of the periapsis in the $p_{\rm orb}$ plane, and $i$ is the inclination\footnote{Here we define inclination as an angle between the normal to orbital plane and the direction toward the observer} ($i=0$ for face-on orbits). The projected orbit has, in general, a different eccentricity and separation than the original one. We calculate $a_{\rm proj}$ as the semi-major axis of the projected orbit. 

\begin{figure}
    \centering
    \includegraphics[width=\columnwidth]{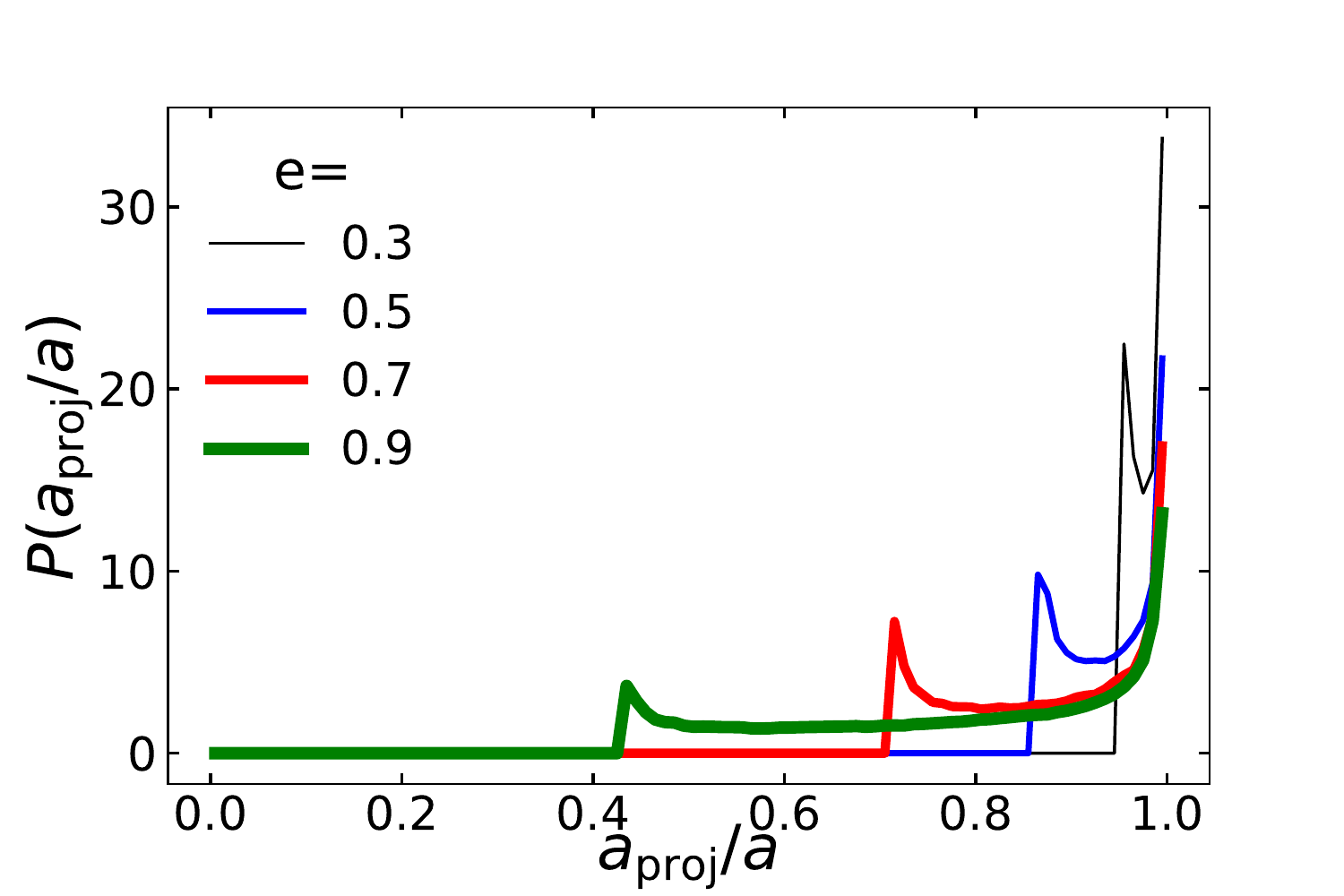}
    \caption{Probability distribution for the ratio between the separation of the projected orbit ($a_{\rm proj}$) and the separation of the original orbit ($a$) for a few example values of eccentricity ($e$). All orientations of the original orbit are assumed to be equally probable.
    }
    \label{fig:Paproj}
\end{figure}

The semi-minor axis of the companion's orbit ($b_{\rm comp}=a_{\rm comp}\sqrt{1-e^2}$) and its semi-major axis ($a_{\rm comp}$) are the minimal/maximal possible values of $a_{\rm proj}$. The probability distribution of $a_{\rm proj}$ ($P(a_{\rm proj})$), which depends on the original orbit's semi-major axis ($a$) and eccentricity ($e$), is calculated numerically (see Figure~\ref{fig:Paproj}). $P(a_{\rm proj})$ is nonzero only for $a_{\rm proj}\in [b_{\rm comp}, a_{\rm comp}]$. We calculate $P(a_{\rm proj})$ numerically using Equation~\ref{eq:r_obs} and assuming that the positional angle of periapsis is distributed uniformly between 0 and $2\pi$, whereas the inclination has a probability distribution of $P(i)=0.5\sin i$. 

Weight $f_{\rm w}$, which is frequently used for the results of PS simulations, is calculated as
\begin{equation}\label{eq:f_w}
    f_{\rm w}=f_{\rm scale}\; f_{\rm SFH} \; f_{\rm MWC}(Z),
\end{equation}
\noindent where $f_{\rm scale}=M_{\rm MW}/M_{\rm sim}$ is the scaling factor from the simulated stellar mass $M_{\rm sim}$ to the mass of the MW \citep[$M_{\rm MW}=\sci{6.08}{10}\Msun$;][]{Licquia1506}. 
$M_{\rm sim} = \sci{4.8}{8}\Msun$ for all the models except \imfs\ and \imff, for which it is  $\sci{1.1}{9}\Msun$ and $\sci{2.4}{8}\Msun$, respectively. $f_{\rm SFH}$ is a probability that the system will be observed currently, i.e. that it was formed $t_{\rm age}$ years ago. In the case of the model used in this study (Section~\ref{sec:MWmodel}), in which the star formation is constant between $t_{\rm SF, start}$ and $t_{\rm SF, end}$ (the values are provided in Table~\ref{tab:MWcomponents} for all MW components), it can be calculated as
\begin{flalign}\label{eq:f_SFH}
    \nonumber &f_{\rm SFH}=\\
    &\frac{\max(0,\min(t_{\rm age}+dt_{\rm age},t_{\rm SF, start})-\max(t_{\rm age},t_{\rm SF,end}))}{t_{\rm SF,start}-t_{\rm SF,end}},
\end{flalign}
\noindent which is the fraction of the stellar mass formed within $[t_{\rm age}, t_{\rm age}+dt_{\rm age}]$, where $dt_{\rm age}$ is the evolutionary time step from PS (the integral over the nBHB phase duration gives $\Delta t_{\rm nBHB}$ used in Equation~\ref{eq:nBHB_Gaia}). For more general approach to calculating $f_{\rm SFH}$ see \citet[Equation~8]{Wiktorowicz1904}. 

$f_{\rm MWC}(Z)$ is a fraction of the MW mass in a Galactic component that has a metallicity $Z$. According to our MW model, the values of $f_{\rm MWC}$ are equal to $M_{\rm tot}/M_{\rm MW}$, where values of $M_{\rm tot}$ are provided in Table~\ref{tab:MWcomponents}, for binaries with metallicity $Z$ equal to the metallicity of the MW component ($Z$ it Table~\ref{tab:MWcomponents}). $f_{\rm MWC}$ is zero for binaries with different metallicity.

Finally, $P(d<d_{\rm max})$ is a probability that a randomly located nBHB (according to spatial distributions in Table~\ref{tab:MWcomponents}) will have a distance to the observer lower than $d_{\rm max}$, which is the maximal distance at which the companion is still visible in \gaia, 
$d_{\rm max}=\min(d_{\rm astro}, d_{\rm photo})$. The limit $d_{\rm astro}$ comes from the fact that the apparent size of the orbit ($a_{\rm proj}$) must be large enough to enable detection by \gaia. The angular size of the orbit is calculated as $\alpha=a_{\rm proj}/d=4.65\microarcsec (a_{\rm proj}/\Rsun) (d/\kpc)^{-1}$, where $d$ is a distance to the binary. We assume that only systems for which $\alpha>\sigma_{\rm G}$, where $\sigma_{\rm G}$ depend on the object's apparent luminosity $m$ \citep{Gaia1611}, will be detectable by \gaia. Consequently, $d_{\rm astro}$ comes as a solution to $d_{\rm astro}\sigma_{\rm G}(m(d_{\rm astro},L))=a_{\rm proj}$, where $L$ is the companion's luminosity and the apparent magnitude $m$ is calculated as
\begin{equation}\label{eq:m}
    m=5\log_{10}\left(\left(\frac{d}{\kpc}\right)\left(\frac{L}{\Lsun}\right)^{-\frac{1}{2}}\right)+d+14.83,
\end{equation}
\noindent where we take into account the interstellar extinction as $A(d)=d/\kpc$ \citep[assuming that the \gaia\ G band is nearly equal to $V$ band]{Spitzer78}. We note that the extinction is expected to saturate at large distances \citep[e.g.][]{Sale0901} and to be smaller for higher galactic latitudes \citep[e.g.][]{Marshall0607}; however, these effects affect mostly the remote parts of the disk and halo, where only a minor fraction of the Galactic stars are located. Therefore, we use here the simple formula and plan to investigate the effects of extinction in more detail in future work. $d_{\rm photo}$ is the photometric distance, i.e. the distance at which the apparent magnitude of the companion becomes equal to the photometric limit of \gaia\ ($m_{\rm lim, \gaia}=21$) and can be calculated as a solution to $m(d_{\rm photo}, L)=m_{\rm lim, \gaia}$.

After we calculate $d_{\rm max}$, the $P(d<d_{\rm max})$ can be calculated assuming that the spatial distribution of nBHBs follows the spatial distribution of stars. Then, this probability is equal to the fraction of the Galactic mass within the sphere centered on the Sun and radius $r=d_{\rm max}$, i.e. $P(d<d_{\rm max})=f_{\rm gal}(d_{\rm max})$ (Figure~\ref{fig:profiles})

\subsubsection{Procedure for \lamost}\label{sec:nBHB_LAMOST}

In the case of \lamost, in our analysis we have included only stars located in the \lamost\ field of view (about $\sim6\%$ of all stars in the Galaxy; Table~\ref{tab:LAMOST_FOV}). As \lamost\ is expected to operate longer than \gaia, we chose the maximal orbital period as $10\yr$ for binaries. As \lamost\ is a spectroscopic survey, the procedure is different than for \gaia; however, the general idea stays the same. We calculate the expected number of detections as

\begin{flalign}\label{eq:nBHB_LAMOST}
\nonumber &N_{\rm nBHB, \lamost}=\\
&\sum_{\rm nBHBs}\int_{\Delta t_{\rm nBHB}}dt P(\Delta RV>RV_{\rm err}) f_{\rm w} P(d\leq d_{\rm photo}) f_{\rm visible}.
\end{flalign}
\noindent Similarly to the \gaia\ procedure (Section~\ref{sec:nBHB_Gaia}), we sum the results for all the binaries in the database, and integration goes over the nBHB phase ($\Delta t_{\rm nBHB}$). $f_{\rm w}$ is calculated in the same way as for \gaia\ (Equation~\ref{eq:f_w}). The other factors are described below.

\begin{figure}
    \centering
    \includegraphics[width=\columnwidth]{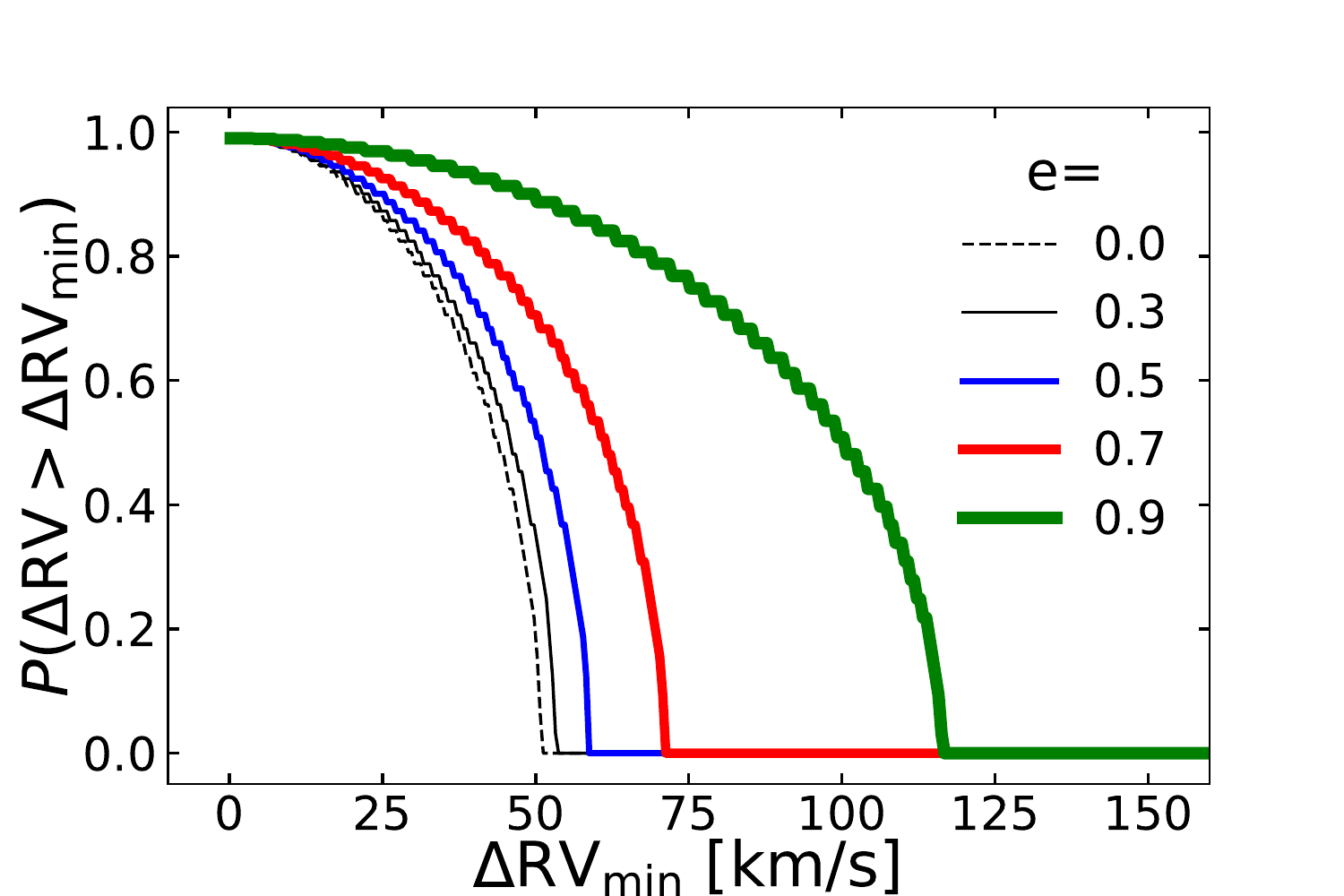}
    \caption{Probability distribution of RV variations ($\Delta \mathrm{RV}$) being larger than the limiting value ($\Delta \mathrm{RV}_{\rm min}$). Provided values of $\Delta \mathrm{RV}_{\rm min}$ are for a reference binary with separation $a=1\Rsun$ and orbital period $P=1$ day. For other binaries the scaling should be applied, $\Delta \mathrm{RV} = \Delta\mathrm{RV}_{\rm obs} (P/\mathrm{day})(a/\Rsun)^{-1}$, where $\Delta\mathrm{RV}_{\rm obs}$ is the observed RV variation for a binary with period $P$ and separation $a$, in order to use the presented relations. Distributions for a few example values of eccentricity ($e$) are provided.}
    \label{fig:Prv}
\end{figure}

The change of RV observable by \lamost\ has to be higher than the RV error, which is equal to $\mathrm{RV}_{\rm err}=10 \kms$ for OB stars and lower for fainter stars (Bai et al., in preparation) We calculate a probability that RV of a randomly oriented orbit is higher than this limit as $P(\Delta\mathrm{RV}>\mathrm{RV}_{\rm err})$. For any orbital orientation, the RV variation ($\Delta\mathrm{RV}$) is calculated as a half of the difference between the highest and lowest RV as observed from the location of the Sun. We assume a random distribution of orbital orientations and calculate this probability distribution numerically (see Figure~\ref{fig:Prv}). Only inclined ($i>0^\circ$) systems can be spectroscopically detectable, but if the orbital velocities are small (e.g. wide circular binaries), the system will not be visible even observed edge-on ($i\approx90^\circ$). On average, the lower the inclination, the lower is $\Delta RV$, however, for the randomly orientated orbits higher inclinations ($i>60^\circ$) are preferred to lower ones ($i<30^\circ$; the probability distribution for the inclination of randomly orientated orbits is $P(i)=0.5 \sin i$). 

$P(d\leq d_{\rm photo})$ is calculated in the same way as $P(d\leq d_{\rm max})$ in the \gaia\ case. The photometric detection limit for \lamost\ is lower than for \gaia\ and equals $m_{\rm lim, \lamost}=16$. We note that as long as the companion is within this magnitude limit, the RV measurements do not depend on distance. Actually, \citet{Deng1207} gave an average $V$-band limiting magnitude of $18$ for \lamost\ low-resolution spectrograph, although the \lamost\ medium-resolution spectrograph has a smaller $V$-band limiting magnitude \citep[$15$\,mag;][]{Liu1905}. However, this limit actually depends on the design of the specific survey. As a parameter study we test also $m_{\rm lim}=20$ (see Table~\ref{tab:results}), which may be reached in future by spectroscopic surveys, like the DESI survey \citep{2016arXiv161100036D} and PFS survey \citep{2014PASJ...66R...1T}. Despite being much deeper, such a survey is expected to increase the detection rates of nBHBs by a factor of only $\sim4$. It results from the fact that the fraction of galaxies observable by \lamost\ ($f_{\rm gal}$; Figure~\ref{fig:profiles}) grows nearly linearly with distance. Equation~\ref{eq:m} gives us the following relation between the apparent magnitude limit and the maximum distance to which a particular star is visible:
\begin{equation}\label{eq:dm}
    m_2-m_1=5\log_{10}\frac{d_2}{d_1}+d_2-d_1.
\end{equation}
For $\Delta m=4$ (increase from $m_{\rm lim}=16$ to $m_{\rm lim}=20$) the distance to which stars are observable increases by a factor of less than $3$ if their luminoisity is $L\gtrsim\Lsun$. For less luminous stars the increase is higher (up to $\sim4$), but these stars are still visible only in the vicinitiy of the Sun ($d\lesssim3\kpc$). As a result, the distance to which the stars are observable grows on average by a factor of $\sim4$ (majority of stars observable by \lamost\ are low mass, see Section~\ref{sec:results}). The number of stars observable by LAMOST grows linearly with distance (Figure~\ref{fig:profiles}); therefore, the predictions grow also by a factor of $\sim4$ (we assume that the spatial distribution of nBHBs follows the spatial distribution of stars). We can also expect that for intermediate values of the limiting apparent magnitude the number of nBHBs observable by \lamost\ will change linearly. It is motivated by the fact that in Equation~\ref{eq:dm} $\Delta d=d_2-d_1$ dominates over the logarithm, so that $\Delta m\approx\Delta d$. For \lamost, $f_{\rm gal}$ grows nearly linearly between $d\approx3\kpc$ and $10\kpc$ (Figure~\ref{fig:profiles}); therefore, $N_{\rm nBHB,\lamost}\propto f_{\rm gal}\propto d\propto m$.

$f_{\rm visible}$ is the fraction of a galactic component visible by \lamost. The values calculated for the \lamost\ field of view are provided in Table~\ref{tab:LAMOST_FOV}.

\section{Results}\label{sec:results}

\begin{deluxetable*}{lllll}
    \tablewidth{\textwidth}
    \tablecaption{Predictions for the Milky Way galaxy}

    \tablehead{ model & $N_{\rm nBHB}$ & $N_{\rm nBHB, Gaia}$ & $N_{\rm nBHB, LAMOST}$ & $N_{\rm nBHB, LAMOST^*}$ ($n_{\rm lim}=20$)}
    \startdata 
        \std       & \sci{1.5\pm0.01}{6} & \sci{1.6\pm0.002}{2} & \sci{1.3\pm0.02}{-1} & \sci{2.8\pm0.05}{-1} \\
        \ssa       & \sci{3.2\pm0.02}{6} & \sci{3.4\pm0.009}{2} & \sci{1.3\pm0.007}{1}  & \sci{5.0\pm0.02}{1} \\
        \nkr       & \sci{1.5\pm0.04}{5} & \sci{7.8\pm0.02}{1} & \sci{1.5\pm0.02}{-1}  & \sci{3.3\pm0.06}{-1} \\
        \nkbe      & \sci{1.5\pm0.04}{5} & \sci{7.7\pm0.02}{1} & \sci{1.7\pm0.02}{-1}  & \sci{3.9\pm0.05}{-1} \\
        \imfs      & \sci{5.0\pm0.05}{5} & \sci{4.1\pm0.006}{1} & \sci{3.8\pm0.09}{-2} & \sci{8.2\pm0.1}{-2} \\
        \imff      & \sci{3.1\pm0.03}{6} & \sci{4.7\pm0.004}{2} & \sci{2.8\pm0.03}{-1}  & \sci{6.1\pm0.09}{-1}      

    \enddata

    \tablecomments{Predictions for the number of nBHBs in the MW ($N_{\rm nBHB}$) for investigated models. $N_{\rm nBHB, Gaia/LAMOST}$ are the nBHBs observable by \gaia/\lamost\ (see Section~\ref{sec:obscut} for the details of the observational cuts). In the case of \lamost, 
    the limiting apparent magnitude is set to $m_{\rm lim}=16$. For comparison, results for the magnitude of a \lamost-like spectroscopic survey with an apparent magnitude limit of $20$ ($N_{\rm nBHB, LAMOST^*}$) are also provided. 
    The uncertainties were calculated on the basis of $100$ bootstrap estimates from a population of $>10,000$ nBHBs observable by \gaia\ or \lamost, which were obtained from PS simulations. For $N_{\rm nBHB}$ Poisson uncertainties are provided. In general, systems observable by \gaia\ represent a different subpopulation of nBHBs than those observable by \lamost\ (see text for details).}
    \label{tab:results}
\end{deluxetable*}

The main results, i.e. the predicted number of detections, for each tested model are summarized in Table~\ref{tab:results}. In general, the number of nBHBs detectable by \gaia\ or \lamost\ constitutes only a small fraction ($\lesssim0.051\%$) of all nBHBs expected to exist currently in the MW. For all models, the fraction is higher for \gaia\ than for \lamost\ by at least one order of magnitude. One of the main reasons is the fact that in the \lamost\ field of view there is only a small fraction of the Galaxy ($\lesssim5$--$7$), whereas \gaia\ observes the entire sky. Additionally, the \gaia\ sample consists mostly of luminous stars, which are visible from large distances, especially from the vicinity of the Galactic bulge. On the other hand, in the \lamost\ sample there are more low-mass stars ($\lesssim2\Msun$), which obtain higher luminosities ($\gtrsim200\Lsun$) only after evolving off the main sequence (MS), thus during a short period of their lives. Consequently, nBHB observed by \lamost\ reside mostly in the vicinity of the Sun, where the star number density is low in relation to the vicinity of the Galactic center. 

The observable fraction by \gaia\ is the highest for models with higher NKs (\nkr\ and \nkbe). For these models, the total number of nBHBs drops by an order of magnitude, whereas the observed numbers of nBHBs drop by half (\gaia) or two orders of magnitude (\lamost). The typical companions are responsible for this behavior. In nBHBs, BHs form typically from the primary stars, i.e. the more massive star in each binary on the zero-age MS (ZAMS). As a result of the assumed flat initial mass ratio distribution \citepalias{Wiktorowicz1911}, lower-/higher-mass secondaries are typically associated with lower-/higher-mass BH progenitors\footnote{The distribution of primary mass \Ma\ for secondary masses constrained to the range $\Mb\in[M_1,M_2]$ can be expressed as a conditional probability $P(A|B)$ where $A$ means the primary mass $\Ma=M$ and $B$ means $\Mb\in[M_1,M_2]$. The probability can be calculated as $P(A|B)=P(A)*P(B|A)/P(B)$, where $P(A)=\mathrm{IMF}(M)$. $P(B|A)$ is equal to 0 if $M\leq M_1$, $(M-M_1)/(M-0.08\Msun)$ if $M_1<M\leq M_2$, and $(M_2-M_1)/(M-0.08\Msun)$ if $M_2<M$. $P(B)=\int_{M_1}^{M_2}\int_{x}^{150\Msun}\mathrm{IMF}(M)/(M-0.08)dM dx$. $P(A|B)$ drops as $M^{\Gamma-1}$ above $M_2$, therefore, if $M_2$ is small, \Ma\ is also small on average.}.
Only low-mass BH progenitors undergo supernova (SN) explosions with small fallback and, therefore, have significant NKs. Consequently, nBHBs, which might be observable by \lamost\ because they have low-mass companions (see Section~\ref{sec:companions}) in the \std\ model, in the \nkr\ and \nkbe\ models are more frequently disrupted during the BH formation process. We note that some systems may also be excited to wider orbits with periods too large or velocities too small to be detectable by \lamost. In contrast, the \gaia\ sample contains mostly nBHBs with massive stars (Section~\ref{sec:companions}), which must be accompanied by BHs formed from massive progenitors either with strong fallback (low effective NK) or through direct collapse (no NK).

The steepness of the IMF changes the ratio of high-mass to low-mass stars on the ZAMS. The flatter is the IMF, the more BH progenitors are present in the initial (ZAMS) populations, and their average mass is higher. The former directly influences the number of nBHBs ($N_{\rm nBHB}$) and, simultaneously, the numbers of nBHBs observable by both \gaia\ and \lamost. In the \imff\ model, in which the IMF is flatter (the IMF exponent $\Gamma=-1.9$) the predicted numbers are higher than in the \std\ model ($\Gamma=-2.3$), whereas in the \imfs\ model ($\Gamma=-2.7$), the numbers are lower (see Table~\ref{tab:results}). As the tested values of $\Gamma$ are rather extreme, we may conclude that the steepness of the IMF influences the numbers by a factor of $2$--$3$ in relation to the fiducial model. The higher average mass changes the relative fraction of massive BH progenitors and, consequently, the number of massive secondaries. As a result, the fraction of nBHBs observable by \gaia\ is higher/lower in the \imff/\imfs\ model ($0.015\%/0.008\%$) than in the reference model ($0.011\%$). The steepness has a smaller effect on the low-mass end of the IMF; therefore, the effect on the fraction observable by \lamost\ is negligible (Table~\ref{tab:results}).

\subsection{Formation Routes}

\begin{deluxetable*}{lll}
    \tablewidth{\textwidth}
    \tablecaption{Formation roots}

    \tablehead{ Model & \gaia & \lamost}
    \startdata        
        \multicolumn{3}{c}{All nBHBs}\\
        \std     & ($0.69$) MT1(1/2-1) SN1  & ($0.63$) CE1(4-1;7/8-1) SN1 MT2(14-1/2/3/4/5/6)  \\ 
                 & ($0.17$) CE1(4-1;7-1) SN1  & ($0.32$) CE1(4-1;7-1) SN1  \\ 
                 & ($0.14$) CE1(4-1;7/8-1) SN1 MT2(14-1/2/3/4/5/6)  &  \\ 
        \ssa     & ($0.47$) CE1(4-0/1;7-0/1) SN1  & ($0.87$) CE1(4-0/1;7-0/1) SN1  \\ 
                 & ($0.34$) MT1(1/2-1) SN1  & ($0.13$) CE1(4/5-0/1;7/8-0/1) SN1 MT2(14-0/1/2/3/4/5/6)  \\ 
                 & ($0.19$) CE1(4/5-0/1;7/8-0/1) SN1 MT2(14-0/1/2/3/4/5/6)  &  \\ 
        \nkr     & ($0.37$) MT1(1/2-1) SN1  & ($0.68$) CE1(4-1;7/8-1) SN1 MT2(14-1/2/3/4/5/6)  \\ 
                 & ($0.33$) CE1(4-1;7/8-1) SN1 MT2(14-1/2/3/4/5/6)  & ($0.20$) CE1(4-1;7-1) SN1  \\ 
                 & ($0.29$) CE1(4-1;7-1) SN1  & ($0.09$) SN1  \\ 
        \nkbe    & ($0.37$) CE1(4-1;7/8-1) SN1 MT2(14-1/2/3/4/5/6)  & ($0.79$) CE1(4-1;7/8-1) SN1 MT2(14-1/2/3/4/5/6)  \\ 
                 & ($0.33$) MT1(1/2-1) SN1  & ($0.20$) CE1(4-1;7-1) SN1  \\ 
                 & ($0.30$) CE1(4-1;7-1) SN1  &  \\ 
        \imfs    & ($0.66$) MT1(1/2-1) SN1  & ($0.60$) CE1(4-1;7-1) SN1 MT2(14-1/2/3/4/5/6)  \\ 
                 & ($0.17$) CE1(4-1;7-1) SN1  & ($0.35$) CE1(4-1;7-1) SN1  \\ 
                 & ($0.16$) CE1(4-1;7-1) SN1 MT2(14-1/2/3/4/5/6)  &  \\ 
        \imff    & ($0.73$) MT1(1/2-1) SN1  & ($0.72$) CE1(4-1;7/8-1) SN1 MT2(14-1/2/3/4/5/6)  \\ 
                 & ($0.14$) CE1(4-1;7-1) SN1  & ($0.17$) CE1(4-1;7-1) SN1  \\ 
                 & ($0.12$) CE1(4-1;7/8-1) SN1 MT2(14-1/2/3/4/5/6)  &  \\ 
        \multicolumn{3}{c}{ nBHBs with massive BHs}\\
        \std     & ($0.96$) CE1(4-1;7-1) SN1 MT2(14-1/2/3/4) MT2(14-8/9)  & ($0.95$) CE1(4-1;7-1) SN1 MT2(14-1/2/3/4) MT2(14-8/9)  \\ 
        \ssa     & ($0.98$) CE1(4-0/1;7-0/1) SN1  & ($0.98$) CE1(4-0/1;7-0/1) SN1  \\ 
        \nkr     & ($0.91$) CE1(4-1;7-1) SN1 MT2(14-1/2/3/4) MT2(14-8/9)  & ($0.88$) CE1(4-1;7-1) SN1 MT2(14-1/2/3/4) MT2(14-8/9)  \\ 
                 &  & ($0.06$) CE1(4-1;7-1) SN1 MT2(14-1/2/3/4/5/6)  \\ 
        \nkbe    & ($0.96$) CE1(4-1;7-1) SN1 MT2(14-1/2/3/4) MT2(14-8/9)  & ($0.96$) CE1(4-1;7-1) SN1 MT2(14-1/2/3/4) MT2(14-8/9)  \\ 
        \imfs    & ($0.96$) CE1(4-1;7-1) SN1 MT2(14-1/2/3/4) MT2(14-8/9)  & ($0.96$) CE1(4-1;7-1) SN1 MT2(14-1/2/3/4) MT2(14-8/9)  \\ 
        \imff    & ($0.97$) CE1(4-1;7-1) SN1 MT2(14-1/2/3/4) MT2(14-8/9)  & ($0.96$) CE1(4-1;7-1) SN1 MT2(14-1/2/3/4) MT2(14-8/9)  \\ 
        \multicolumn{3}{c}{nBHBs with Massive Companions}\\
        \std     & ($0.99$) MT1(1/2-1) SN1  & ($1.00$) MT1(1/2-1) SN1  \\ 
        \ssa     & ($1.00$) MT1(1/2-1) SN1  & ($1.00$) MT1(1/2-1) SN1  \\ 
        \nkr     & ($0.99$) MT1(1/2-1) SN1  & ($0.99$) MT1(1/2-1) SN1  \\ 
        \nkbe    & ($0.99$) MT1(1/2-1) SN1  & ($0.99$) MT1(1/2-1) SN1  \\ 
        \imfs    & ($0.99$) MT1(1/2-1) SN1  & ($1.00$) MT1(1/2-1) SN1  \\ 
        \imff    & ($0.99$) MT1(1/2-1) SN1  & ($1.00$) MT1(1/2-1) SN1  \\ 
        \multicolumn{3}{c}{Double Compact Objects}\\
        \std     & ($0.59$) MT1(1/2-1) SN1  & ($0.56$) MT1(1/2-1) SN1  \\ 
                 & ($0.32$) CE1(4-1;7-1) SN1  & ($0.21$) CE1(4-1;7-1) SN1  \\ 
                 & ($0.09$) CE1(4-1;7-1) SN1 MT2(14-1/2/3)  & ($0.14$) CE1(4-1;7-1) SN1 MT2(14-2/3)  \\ 
        \ssa     & ($0.68$) MT1(1/2-1) SN1  & ($0.44$) MT1(1/2-1) SN1  \\ 
                 & ($0.25$) CE1(4-1;7-1) SN1  & ($0.24$) SN1 MT2(14-2/3/4)  \\ 
                 & ($0.07$) CE1(4-1;7-1) SN1 MT2(14-1/2/3)  & ($0.13$) CE1(4-1;7-1) SN1  \\ 
                 &  & ($0.10$) CE1(4-1;7-1) SN1 MT2(14-1/2/3/4)  \\ 
                 &  & ($0.06$) SN1  \\ 
        \nkr     & ($0.76$) CE1(4-1;7-1) SN1  & ($0.91$) SN1  \\ 
                 & ($0.17$) CE1(4-1;7-1) SN1 MT2(14-2/3)  &  \\ 
        \nkbe    & ($0.76$) CE1(4-1;7-1) SN1  & ($0.50$) CE1(4-1;7-1) SN1 MT2(14-1/2/3/4)  \\ 
                 & ($0.16$) CE1(4-1;7-1) SN1 MT2(14-1/2/3)  & ($0.25$) MT1(1/2-1) SN1  \\ 
                 & ($0.07$) MT1(1/2-1) SN1  & ($0.24$) CE1(4-1;7-1) SN1  \\ 
        \imfs    & ($0.58$) MT1(1/2-1) SN1  & ($0.64$) MT1(1/2-1) SN1  \\ 
                 & ($0.33$) CE1(4-1;7-1) SN1  & ($0.19$) CE1(4-1;7-1) SN1 MT2(14-2/3)  \\ 
                 & ($0.10$) CE1(4-1;7-1) SN1 MT2(14-2/3)  & ($0.06$) SN1  \\ 
                 &  & ($0.05$) CE1(4-1;7-1) SN1  \\ 
        \imff    & ($0.65$) MT1(1/2-1) SN1  & ($0.60$) MT1(1/2-1) SN1  \\ 
                 & ($0.28$) CE1(4-1;7-1) SN1  & ($0.18$) CE1(4-1;7-1) SN1 MT2(14-1/2/3/4)  \\ 
                 & ($0.07$) CE1(4-1;7-1) SN1 MT2(14-1/2/3/4)  & ($0.14$) CE1(4-1;7-1) SN1
    \enddata
    \tablecomments{Symbolical representations of the typical evolutionary roots leading to the formation of nBHBs that are predicted to be observed by \gaia\ and \lamost. Only main evolutionary phases are presented. The routes for the total samples are presented at the top. Additionally, the typical routes for nBHBs with \emph{massive BHs} ($\Mbh>20\Msun$), \emph{massive companions} ($\Mcomp>30\Msun$), and nBHBs that are \emph{DCO progenitors} are displayed separately. The numbers in parentheses represent the fraction of nBHBs from a particular subgroup that were formed through this route. The symbols represent the following: SN1--supernova of the primary; MT1/2--MT (primary/secondary is a donor); CE1/2(a1-b1;a2-b2)--common envelope (primary/secondary is a donor; a1/2, primary's evolutionary type before/after the CE; b1/2, secondary's evolutionary type before/after the CE). Evolutionary types (numbers inside parentheses) are as follows: 1--MS; 2--Hertzsprung gap; 3--red giant; 4--core helium burning; 5--early asymptotic giant branch; 6--thermal pulsing asymptotic giant branch; 7--helium MS; 8--helium Hertzsprung gap; 9--helium red giant; 13--neutron star; 14--BH.} 
    \label{tab:evroutes}
\end{deluxetable*}

Most of the nBHBs in the Galaxy ($\sim95$--$96\%$) were formed without any strong interactions (MT or CE) in the majority of the tested models. The exception are models with increased average NKs (\nkr\ and \nkbe) in which nBHB progenitors are typically disrupted after compact object formation unless some prior interactions harden the system. Consequently, in these models only $\lesssim5\%$ of systems with no interaction history survive to the nBHB phase. On the other hand, nBHBs in the \gaia\ and \lamost\ samples did undergo an MT or a CE phase (Table~\ref{tab:evroutes}) prior to the BH formation irrespective of the model. We assumed that the limit on the survey duration is simultaneously a limit for the orbital period of the observable astrometric or spectroscopic binary  ($P_{\rm orb}<5\yr$ for \gaia\ or $10\yr$ for \lamost). We note that binaries with even longer orbital periods can also be potentially detected, but in this study we assumed a conservative limit. These orbital periods translate into a limit on separation of $\sim3000$--$4000\Rsun$. For shorter orbits with separations $\lesssim1000\Rsun$, BH progenitors, which expand to at least $1000\Rsun$ if evolving as single stars, easily fill their RL and interact with companions through MT or CE. We note that the MT in binaries with mass ratio $q\approx1$ generally leads to mass ratio reversal and effectively to widening of the orbit. Wider systems with initial separation $a_{\rm ZAMS}\approx1000$--$4000\Rsun$ and mass ratio $q\approx1$ (typical for the \gaia\ sample) also interact, because the RL of the primary is typically $\sim0.4\times a\lesssim1000\Rsun$. In such a case, the CE can be very efficient in decreasing orbital separation. If the MT is stable, usually another phase of MT is necessary after the formation of the BH in order to obtain a binary that can be observed astrometrically by \gaia\ (or \lamost). In the \lamost\ sample, the companions are typically lighter; therefore, the mass ratio is smaller and the primary's RL larger ($\lesssim0.7\times a\lesssim3000\Rsun$). In such cases, the primary may not fill the RL during its evolution and the system may stay wide. However, stars on wide orbits have rather slow orbital velocities and, if they have low masses, are visible only in the vicinity of the Sun, which significantly lowers their detection probability by \lamost, and thus the expected number of detections. As a consequence, the separations of nBHBs in the \lamost\ sample are on average smaller than in the \gaia\ sample. We note that in the \lamost\ sample nBHBs are also present with massive companions, which makes it possible for the primary to fill its RL on a wide orbit (see Section~\ref{sec:companions}). Systems that have initial orbital periods larger than the assumed detection limits in the absence of interactions tend to expand owing to the loss of orbital angular momentum in stellar wind and so become unobservable by \gaia\ or \lamost. Furthermore, such systems, in the lack of interactions, are not expected to become observable by \gaia\ or \lamost. For separations higher than $\sim4000\Rsun$, RLs are generally too large to be filled by typical BH progenitors unless the orbit is significantly eccentric. Then, the RL may be filled in periapsis, where the tidal interactions may circularize the system, and thus significantly lower the initially large separation. Although in the \std\ model initial eccentricities are typically low, the \ssa\ model assumes an initial distribution of eccentricities that is skewed toward higher values ($P(e)\propto e$). Therefore, in this model, systems with high initial separations (up to $\sim10^5\Rsun$) form nBHBs observable by \gaia\ or \lamost\ more frequently than in other tested models, although they have the same initial distribution of masses as the \std\ model.  

\begin{figure*}
    \centering
    \includegraphics[scale=0.43]{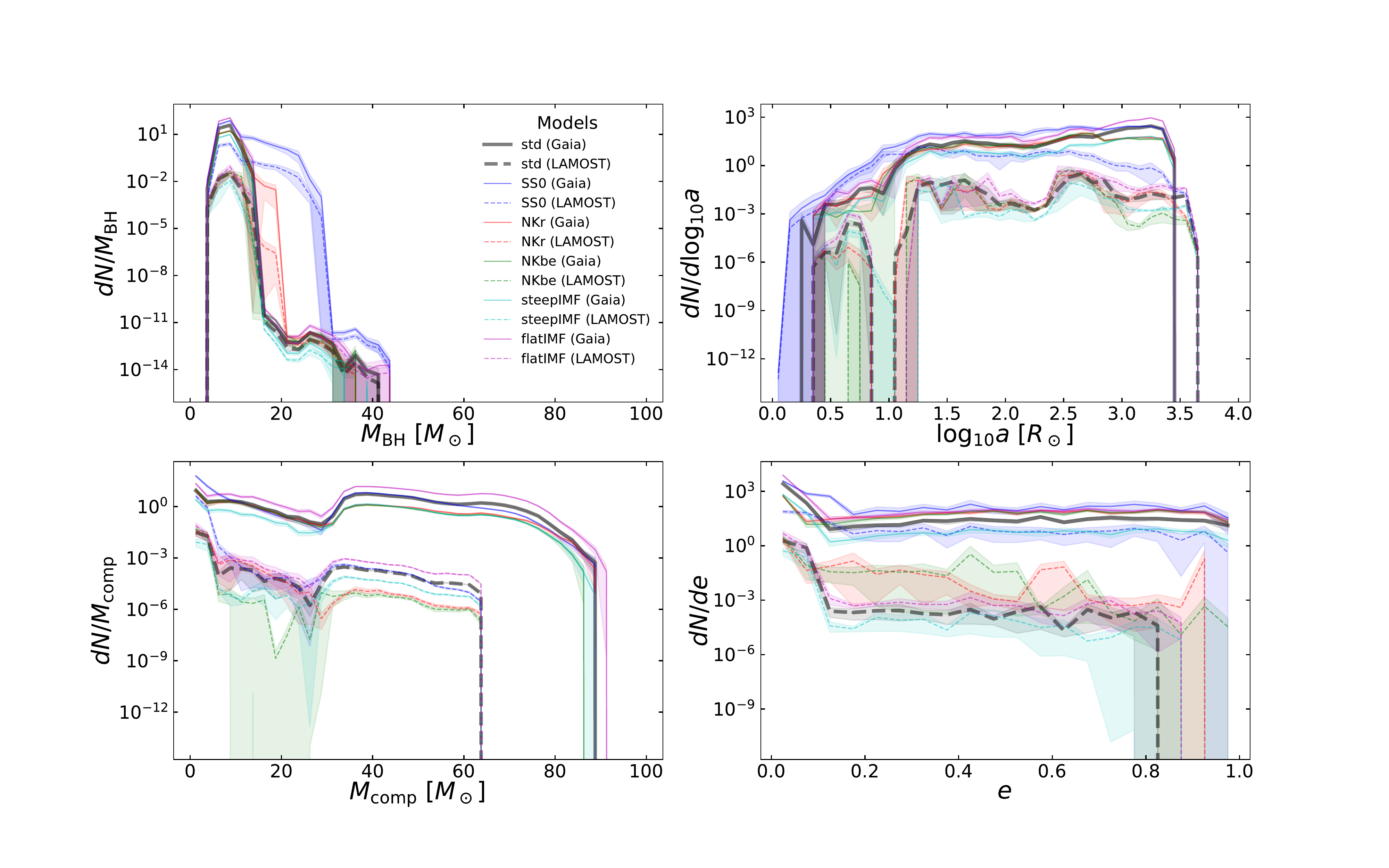}
    \caption{Parameter distributions for nBHBs observable by \gaia\ and \lamost. A comparison of tested models is provided (see Section~\ref{sec:binary_evolution}). $M_{\rm BH/comp}$--BH/companion mass; $a$--separation; $e$--eccentricity; $N$--number. \new{Bands represents $95\%$ confidence limits calculated as such $[x_1, x_2]$ such that $x_1 -x_2 = \min\{x'_1 - x'_2: \sum_{x'\in[x'_1, x'_2]} w_{x'} / \sum_x w_{x'} \geq 0.95\}$ where $x'$ and $w_{x'}$ are data points and their respective weights.}}
    \label{fig:dists}
\end{figure*}

\begin{figure*}
    \centering
    \includegraphics[scale=0.43]{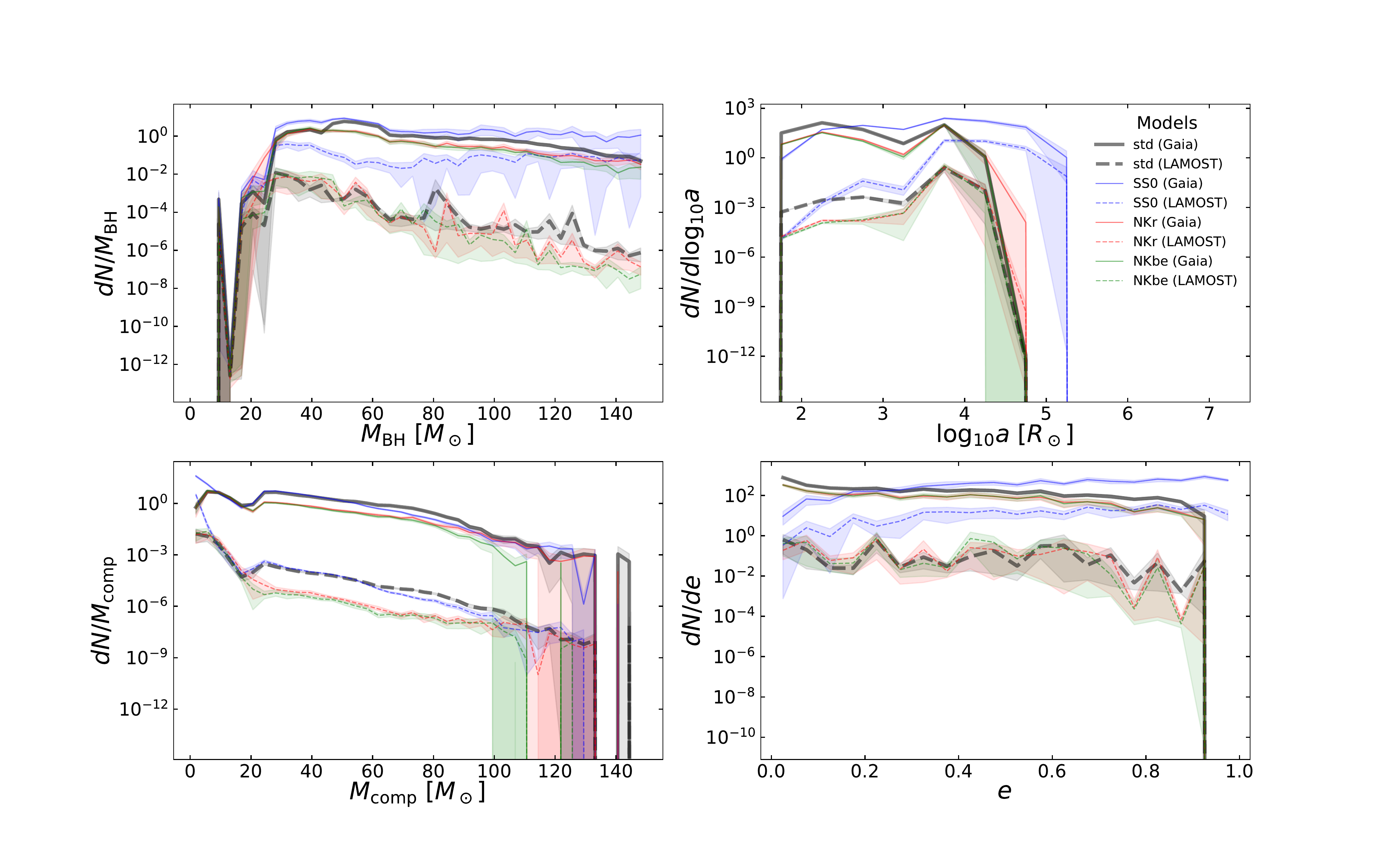}
    \caption{Similar to Figure~\ref{fig:dists}, but for initial (ZAMS) populations.} 
    \label{fig:dists_ZAMS}
\end{figure*}

\begin{figure}
    \centering
    \includegraphics[width=\columnwidth]{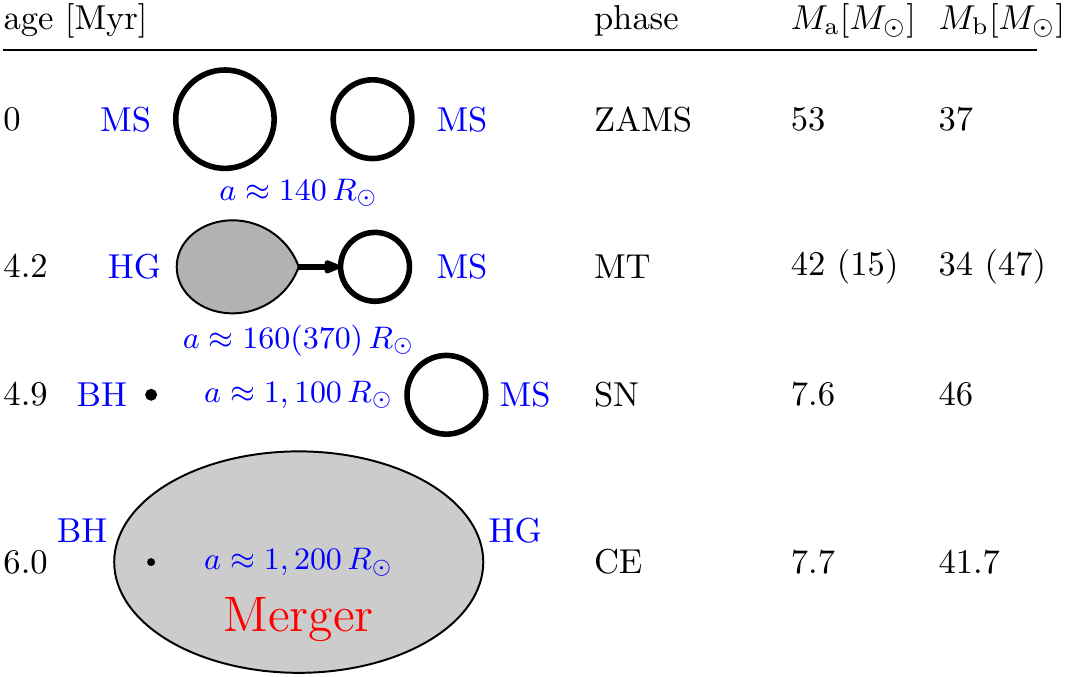}
    \caption{Example of a typical system evolution leading to the formation of an nBHB detectable by \gaia. The phases are as follows: ZAMS--zero age MS; MT--mass transfer; SN--supernova, i.e. formation of the BH; CE--common envelope. Stellar evolutionary phases are as follows: MS--main sequence; HG--Hertzsprung gap;  BH--black hole. Numbers in parentheses express a change of the parameter during the phase. The binary on the ZAMS is composed of a $\sim53\Msun$ primary and a $\sim37\Msun$ secondary on a relatively short orbit of $140\Rsun$. After about $4.2\myr$, the primary evolves off the MS and fills its RL while expanding on the HG. The MT commences, which within $1.4\kyr$ reduces the primary mass to $15\myr$. Half of this mass is acquired by the secondary, the rest being rejected from the system. The primary evolves into a helium star and within $700\kyr$ forms a BH with a small NK. The nBHB forms with an MS companion and a separation of $a\approx1100\Rsun$ which lasts for $\sim1.1\myr$. Afterward, the secondary evolves off the MS and fills its RL while expanding on HG. This time the mass ratio is higher, $q\approx6$, which makes the MT unstable, and the CE occurs. The primary cannot reject the envelope, which results in a merger.}
    \label{fig:ts_Gaia}
\end{figure}

Typical systems in a \gaia\ sample have a BH mass of about $7$--$8\Msun$ and a secondary mass of $35$--$50\Msun$, whereas the typical separation is of the order of $1000$--$2000\Rsun$ (Figure~\ref{fig:dists}). Such nBHBs originate from initial binaries with $\Mzamsa=45$--$55\Msun$, $\Mzamsb=25$--$40\Msun$, and $a_{\rm ZAMS}=100$--$500\Rsun$ (Figure~\ref{fig:dists_ZAMS}). In the typical scenario (Figure~\ref{fig:ts_Gaia}), the massive primary fills its RL while expanding on the Hertzsprung gap (HG), or even already on the MS, and transfers mass to the secondary. Meanwhile, the mass ratio reverses and the orbit starts to expand. After the primary forms a BH, the nBHB is observable by \gaia\ for $1$--$2.5\myr$ before the secondary fills its RL owing to evolutionary expansion. The further evolution can be twofold: (1) In the typical case the MT is unstable due to the high mass ratio and results in a CE and, typically, a merger. In other cases, (2) the MT is stable and the donor loses its hydrogen envelope, becoming a helium star. Afterward, the system may be observable as an nBHB with a helium star companion (Figure~\ref{fig:Kdist}), whose lifetime is, however, much shorter ($\lesssim0.5\myr$) than in the case of an nBHB with an MS companion.

\begin{figure}
    \centering
    \includegraphics[width=\columnwidth]{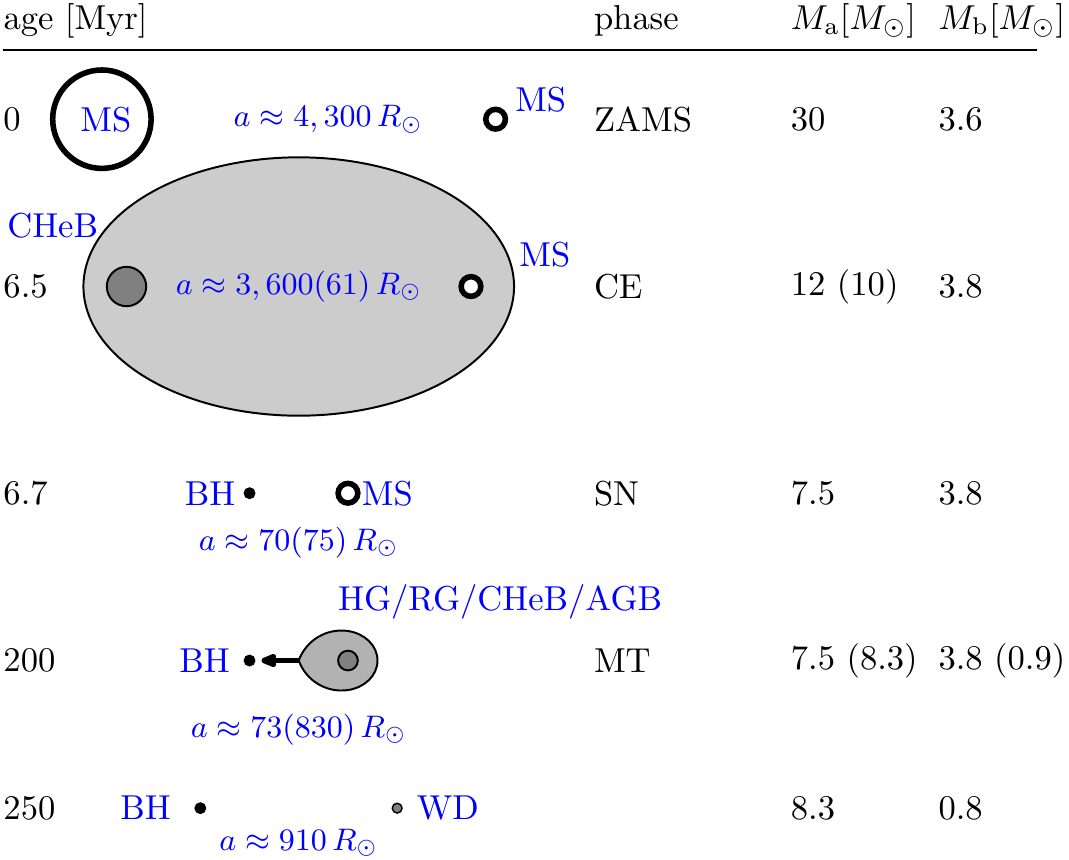}
    \caption{Similar to Figure~\ref{fig:ts_Gaia}, but for typical nBHBs in the \lamost\ sample. The additional evolutionary phases (not explained in the caption to Figure~\ref{fig:ts_Gaia}) are as follows: RG--red giant; CHeB--core helium burning; AGB--asymptotic giant branch; WD--white dwarf. The system begins its evolution as a $30\Msun$ and $3.6\Msun$ binary on a relatively large orbit of $\sim4300\Rsun$. The primary evolves much faster and after $6.5\myr$ fills its RL while being on CHeB. The primary is $12\Msun$ at that moment, whereas the secondary is $3.8\Msun$, therefore, CE commences owing to the high mass ratio. The binary survives, while the separation decreases to $70\Rsun$. Shortly after, the primary, deprived of its hydrogen envelope, forms a $7.5\Msun$ BH with a small NK. An nBHB observable by \lamost\ emerges and exists for nearly $200\myr$. Then, the secondary, while expanding on HG, fills it RL and a stable MT starts, which prolongs for $\sim50\myr$. Meanwhile, the secondary evolves and finally becomes a WD, which ends the system evolution. Although being an nBHB, the separation is too high ($a\approx910\Rsun$) and the WD too dim to be detectable by \lamost.}
    \label{fig:ts_LAMOST}
\end{figure}

In the case of \lamost, the typical evolution of a progenitor of a detectable nBHB looks different (Figure~\ref{fig:ts_LAMOST}) than in the \gaia\ case. The progenitors have masses of $\Mzamsa\approx30\Msun$ and $\Mzams\approx3$--$4\Msun$ separated by $a\approx4000$--$5000\Rsun$ on slightly eccentric orbits $e\lesssim0.2$ (Figure~\ref{fig:dists_ZAMS}). When the primary fills its RL while being in the CHeB phase, the mass ratio is low ($q<1/4$), and thus the MT is unstable and a CE phase occurs. The orbital energy of such a wide binary is large enough to eject the envelope of the massive primary and the binary survives as a much closer one ($a\lesssim100\Rsun$). After the formation of the BH, the system can be detected as an nBHB. Typical BH masses are $7.0$--$8.5\Msun$, whereas companions are typically lighter than $4\Msun$. The separation is typically between $\sim10$ and $100\Rsun$ (Figure~\ref{fig:dists}). After $100\myr$, the secondary evolves off the MS, expands as an HG star, and fills its RL. Then, MT occurs, which widens the system to $a\approx  1000\Rsun$ and results in the loss of the secondary's hydrogen envelope. Shortly after ($\lesssim300\kyr$), it becomes a CO white dwarf (WD). On such a wide orbit a WD with its low luminosity would only be detected in a small volume. 

\subsection{Black holes}

The mass distributions of BHs in nBHBs detectable by \gaia\ and \lamost\ are presented in the top-left panel of Figure~\ref{fig:dists}. The majority of these BHs have masses in the range of $5$--$15\Msun$, which is typical for a solar-metallicity environment. Stars with high initial metallicity undergo a strong mass loss in stellar wind, and in the absence of interactions, masses of BHs are limited to about $15\Msun$. MT from the companion may increase the final BH mass up to $\sim22\Msun$ \citepalias{Wiktorowicz1911}, but configurations allowing for such efficient accretion are very rare in initial populations. Heavier BHs form in lower-metallicity environments. According to our model of the Galaxy, such conditions occur only in the thick disk and halo. These Galactic components constitute only a small fraction ($\sim10\%$) of the MW's mass, which is one of the reasons why there are so few expected discoveries of heavy ($\gtrsim20\Msun$) BHs in all tested models. In reality, the younger thin disk is expected to have a lower metallicity than assumed in our MW model, which later steadily increases to the current value \citep[see, e.g.,][and references therein]{Olejak1908}. Nonetheless, such a correction is expected to have a minimal effect on the \gaia\ sample, which is dominated by massive companions, which must have formed recently when the metallicity in the thin disk was already nearly solar. On the other hand, in the \lamost\ sample, most of the stars are low mass and may potentially have formed several gigayears ago in a lower-metallicity environment. However, as already said, low-mass secondaries tend to have low-mass primaries that are about to form low-mass BHs, no matter what the metallicity is. Therefore, the effect of metallicity evolution in the thin disk on the distribution of BH masses in nBHBs likely to be observed by \gaia\ or \lamost\ is expected to be small.

BH masses in the \lamost\ sample tend to have slightly higher values. Companions in the \lamost\ sample are generally lighter than in the \gaia\ sample; thus, they may potentially originate from older and less metal-rich parts of the Galaxy. Masses of BHs formed in a lower-metallicity environment are expected to be higher on average, therefore pushing the distribution of BH masses toward larger values. However, the fraction of the Galactic stellar mass that has lower metallicity than solar is small; thus, the effect on the typical BH mass is negligible. Except for this feature, the distribution of BH masses observable by \lamost\ is similar to that of  \gaia, as most of the nBHBs evolve without interactions and originate from the same Galactic component, the thin disk, which composes most ($\sim77\%$) of the Galactic stellar mass.

The upper BH mass limit in the distributions comes from pair-instability SNe \citep[e.g.][]{Heger0203} and pulsation pair-instability SNe \citep[e.g.][]{Barkat6703,Woosley1702}. The limit is not strict and depends on modeling \citep[e.g.][]{Farmer1912,Leung1912,Mapelli2001,Renzo2004}. For example, \citet{Belczynski2004} presented models limiting the maximal BH mass from single-star evolution to values between $45$ and $56\Msun$. Although mass accretion may increase the BH's mass after its formation, it is by no more than a few \Msun\ \citepalias{Wiktorowicz1911}. Stellar mergers may potentially fill the gap \citep[e.g.][]{DiCarlo1911}. In the absence of strong interactions, pair-instability (pulsation) SNe occur only for stars born with lower ($\lesssim0.1\Zsun$) metallicity, where the helium cores can grow to masses above $\sim30\Msun$ \citep[e.g.][]{Woosley1702}, i.e., in the thick disk and the bulge as far as our MW model is concerned. We note that \citet{Liu1911} recently discovered a $\sim70\Msun$ BH in the MW. Assuming that this observation is correct \citep[see, e.g.,][]{El-Badry2003}, such a massive BH in an nBHB may be in tension with our results. There are three possible explanations: (1) as the authors say, the BH may actually be a very close inner pair or triple in a hierarchical system; (2) the BH may be an outcome of an earlier merger that either happened in a hierarchical system or acquired a companion later on; or (3) the winds assumed in the simulations of \citetalias{Wiktorowicz1911} were too strong to produce heavy BHs like LB-1 \citep{Belczynski1911}. The first two cases are not subjects of binary PS \citep[see, e.g.,][]{Toonen1701}. The third case, i.e. the influence of stellar winds, will be a subject of our future studies.

An interesting behavior is presented by model \ssa. Although the majority of nBHBs in the \ssa\ model have BH masses below $15\Msun$, similarly to other models, the fraction of systems with massive BHs ($\Mbh>20\Msun$) is much higher ($\sim1.1\%$ in comparison to $\lesssim10^{-11}$; Table~\ref{tab:massiveBH}). Model \ssa\ differs from the reference model only in the initial distributions of separations and eccentricities. Specifically, initial orbits of binaries in the \ssa\ model are more eccentric, which has significant consequences. Firstly, if the orbit is eccentric, it is much easier for a BH progenitor to fill its RL, which is much smaller when the secondary passes through periapsis. Resulting tidal interactions reduce the separation, which can be initially large (up to $10^5\Rsun$), to observable levels $\lesssim3000$--$4000\Rsun$. Therefore, the parameter space in initial distributions for nBHB progenitors in the \ssa\ model also includes wide systems that are not present or very rare in other models. Second, the RLs of BH progenitors in wide but eccentric orbits are small enough to be filled even if the companion is a low-mass star. If the orbit is circular, the primary's RL will be too large for RLOF to occur owing to the small mass ratio. This significantly increases the fraction of nBHBs originating from the old stellar environment, as only low-mass stars can be companions in to such nBHBs. This old lower-metallicty Galactic component may constitute $\sim10\%$ of the stellar mass in the Galaxy. Consequently, these effects increase the number of nBHBs predicted by the \ssa\ model to be detected by both \gaia\ and \lamost. 

\begin{deluxetable}{ccc}
    \tablewidth{\columnwidth}
    \tablecaption{Massive BHs in nBHBs}

    \tablehead{ Model & $N_{\rm nBHB, Gaia}$ & $N_{\rm nBHB, LAMOST}$}
    \startdata        
        \std     & \sci{1.3}{-11} (\sci{7.8}{-12}\%) & \sci{4.8}{-12} (\sci{3.6}{-9}\%) \\
        \ssa     & 3.8 \hfill($1.1\%$) & \sci{1.5}{-1}\hfill ($1.2\%$) \\
        \nkr     & \sci{1.5}{-11} (\sci{1.9}{-11}\%) & \sci{5.9}{-12} (\sci{3.9}{-9}\%) \\
        \nkbe    & \sci{1.5}{-11} (\sci{2.0}{-11}\%) & \sci{5.8}{-12} (\sci{3.4}{-9}\%) \\
        \imfs    & \sci{2.3}{-12} (\sci{5.7}{-12}\%) & \sci{8.8}{-13} (\sci{2.3}{-9}\%) \\
        \imff    & \sci{3.5}{-11} (\sci{7.5}{-12}\%) & \sci{1.3}{-11} (\sci{4.6}{-9}\%) 
    \enddata
    \tablecomments{Number of nBHBs in \gaia/\lamost\ samples that harbor massive BHs ($\Mbh>20\Msun$). Numbers in parentheses represent the fractions of the total sample.}
    \label{tab:massiveBH}
\end{deluxetable}

\begin{figure*}
    \centering
    \includegraphics[width=\textwidth]{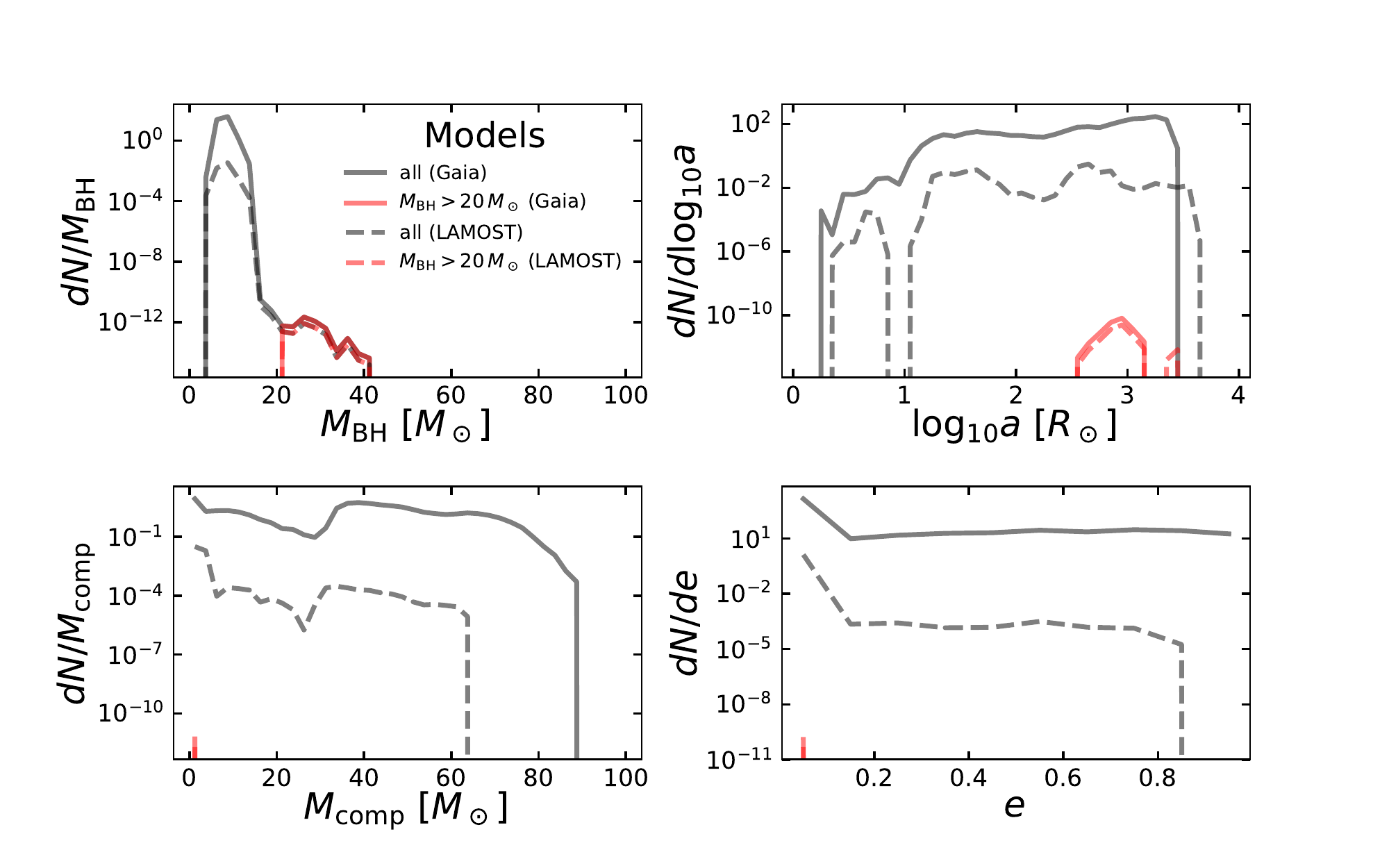}
    \caption{Same as Figure~\ref{fig:dists}, but for the \std\ model only and a subpopulation with massive BHs ($\Mbh>20\Msun$).}
    \label{fig:dists_BH}
\end{figure*}

In general, massive BHs ($\Mbh>20\Msun$) can be found among nBHBs detectable by \gaia\ and \lamost\ in all models, but their fraction among the total population is expected to be, in general, negligibly small (Table~\ref{tab:massiveBH}). These BHs originate from parts of the Galaxy with lower metallicity, i.e. the thick disk and halo. Massive progenitors typically have massive companions; therefore, interactions are inevitable if the final orbital period is to be smaller than $5$ or $10\yr$. In a typical case, the primary fills its RL while expanding during the CHeB phase, which results in a CE. If the system survives the CE, the separation is already significantly smaller. The interactions occur also after the formation of a BH when the secondary expands owing to nuclear evolution and fills its RL. Such a binary might represent an isotropic ultraluminous X-ray sources (ULX) existing in the early MW galaxy \citep[e.g.][]{Wiktorowicz1904}. However, the thick disk and the halo are old systems; therefore, companions to massive BHs in nBHBs either are low-mass stars ($\Mzams\lesssim1\Msun$) or have already evolved into WDs (see Figure~\ref{fig:dists_BH}, bottom left panel). Due to the fact that massive stars had to be accommodated, orbits are relatively large ($a\gtrsim200\Rsun$) and circularized owing to interactions (Figure~\ref{fig:dists_BH}, right panels). 
The number of massive BHs in \gaia\ and \lamost\ samples is small because the birthplaces constrain only a small fraction of the Galaxy ($\sim10\%$) and the companions are low luminous and therefore observable only in the vicinitiy of the Sun ($d\lesssim3\kpc$).

\subsection{Companions}\label{sec:companions}

\begin{figure*}
    \centering
    \includegraphics[width=\textwidth]{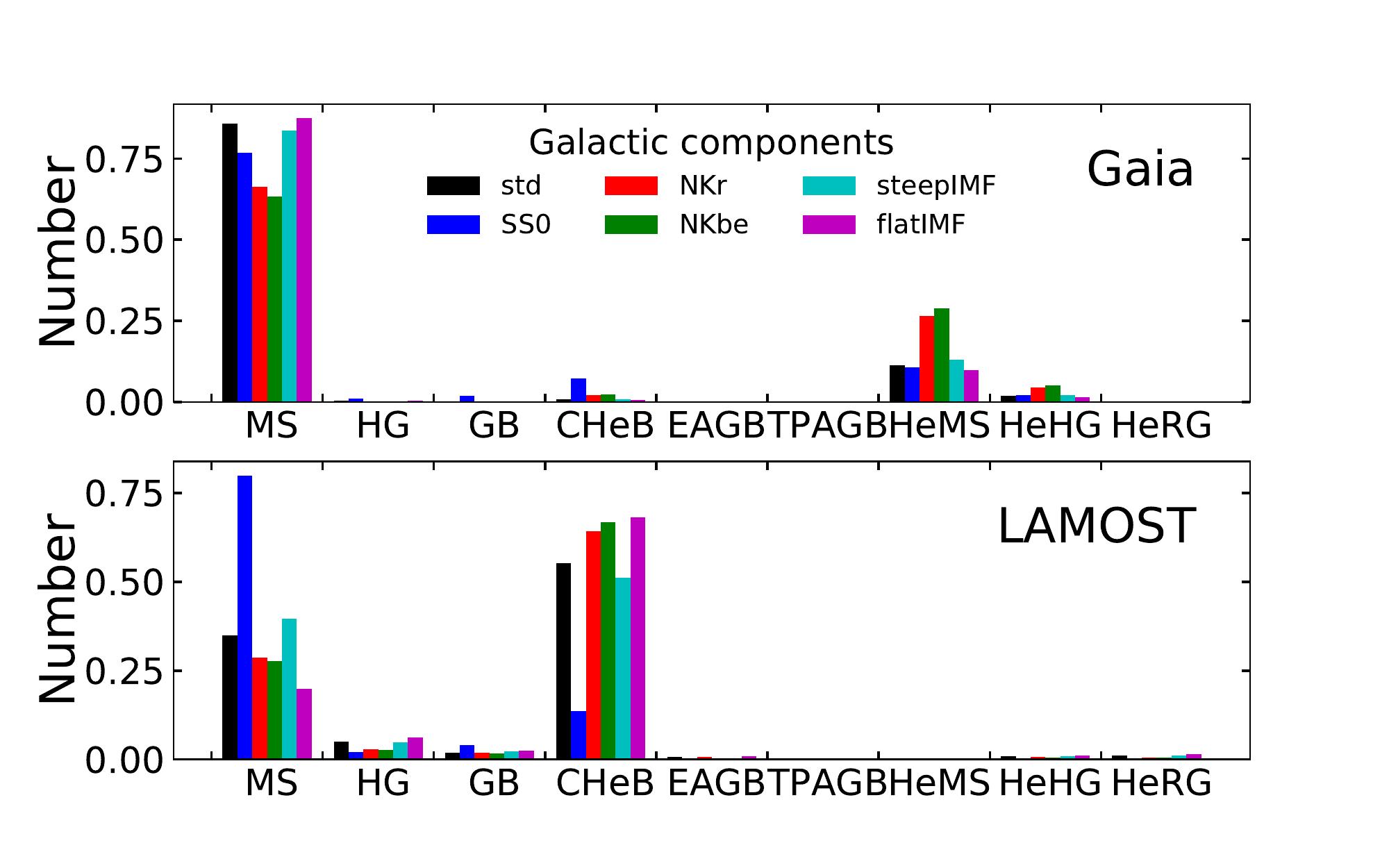}
    \caption{Relative fractions of evolutionary types of companions in nBHBs with division on \gaia\ (top) and \lamost\ (bottom) samples. Both populations are dominated by MS companions in any models, but the \gaia\ sample has typically a significant fraction of helium star companions (HeMS), whereas in \lamost\ giant companions (CHeB) are more typical. The symbols represent the following: MS--main sequence; HG--Hertzsprung gap; GB--giant branch; CHeB--core helium burning; EAGB--early asymptotic giant branch; TPAGB--thermally pulsing asymptotic giant branch; HeMS--helium main sequence; HeHG--helium Hertzsprung gap; HeRG--helium red giant.}
    \label{fig:Kdist}
\end{figure*}

Although BH mass distributions for \gaia\ and \lamost\ samples are very similar, the distributions of companion masses show a significant difference. Particularly, the distributions of companions' evolutionary type vary noticeably (Figure~\ref{fig:Kdist}).

Companions in the \gaia\ sample are mainly OB stars\footnote{For the purpose of this study, which does not involve spectroscopic analysis, we define an OB star as an MS star with a mass above $M>2.1\Msun$ and effective temperature $T>11$ kK.}. Such stars have very short lifetimes ($\lesssim300\myr$) and so can be observed only in environments with recent star formation like thin disks or bulges. Such short lifetimes limit the probability of observation, i.e. the probability that the system was born at such a look-back time that it is observable currently (Equation~\ref{eq:f_SFH}). Only a small fraction ($\sim3\%$) of the thin-disk stellar mass has been formed recently enough to provide OB stars observable currently. On the other hand, among MS stars only OB stars obtain luminosities high enough ($M_{\rm bol}<-2.7$) to be be easily observed by \gaia\ (apparent magnitude limit of $m_{\rm lim}\approx21$) from the vicinity of the Galactic center ($D\approx8.3\kpc$), where most of the stellar mass is spatially localized ($\gtrsim60\%$ of the young thin disk is farther away than $7\kpc$ from the Sun; Figure~\ref{fig:profiles}).

\begin{figure*}
    \centering
    \includegraphics[width=\textwidth]{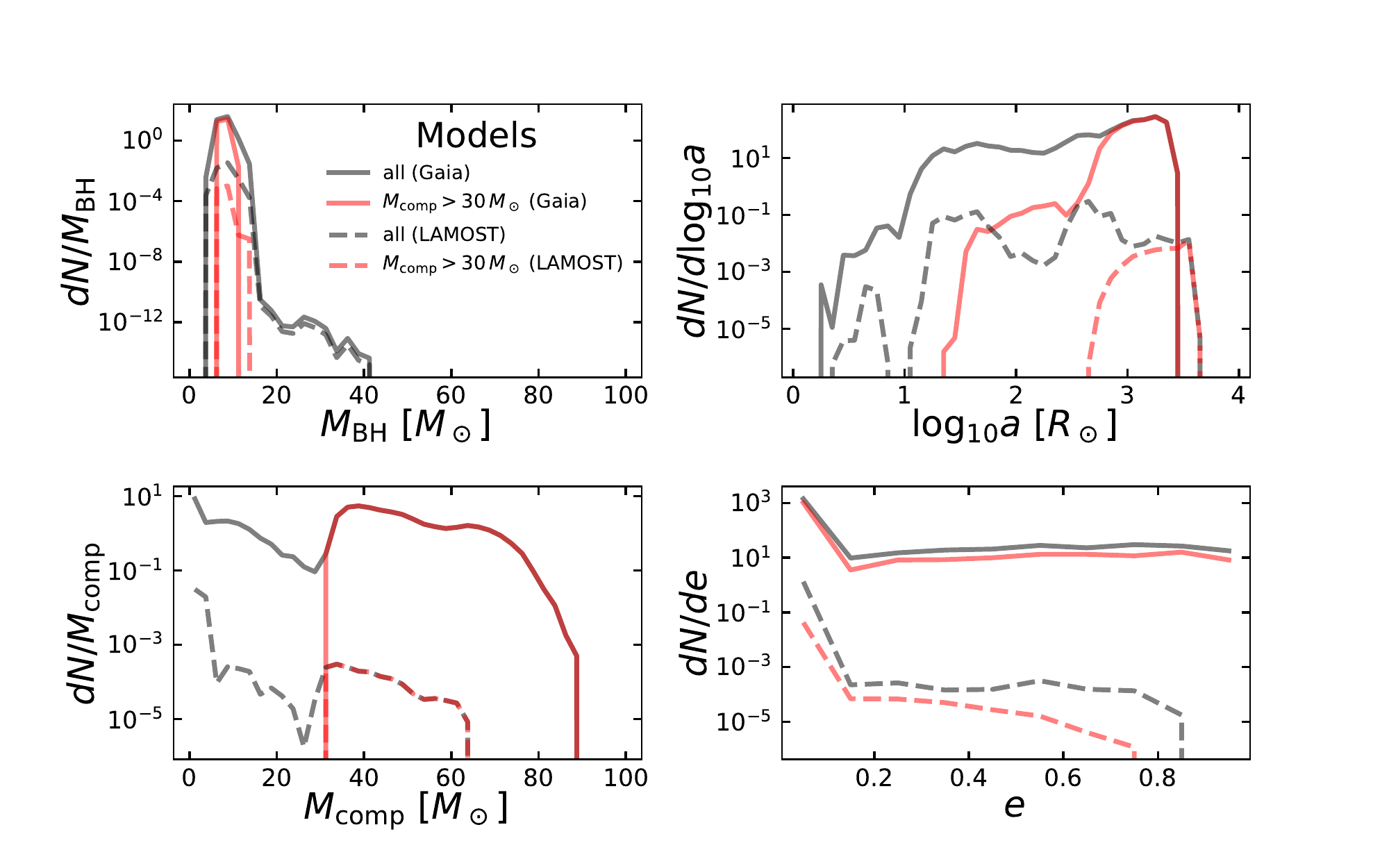}
    \caption{Same as Figure~\ref{fig:dists}, but for the \std\ model only and a subpopulation with massive companions ($\Mcomp>30\Msun$).}
    \label{fig:dists_COMP}
\end{figure*}

A large fraction of detectable companions have high masses \citep[$>30\Msun$;][]{Ziolkowski7201,Langer1209}. Such stars must have formed quite recently ($\lesssim10\myr$) and thus are expected to have rather high metallicities. Therefore, these stars were even heavier on the ZAMS and might have lost a significant fraction of their initial mass in stellar wind. As a consequence, the primaries were also very massive. Such stars expand significantly during their evolution (up to \few{1000\Rsun}) and fill their RLs, which, due to the limitation put on the $P_{\rm orb}$, are $\lesssim1000\Rsun$. During the resulting MT phase, the primary is typically on the HG or still on its MS (Table~\ref{tab:evroutes}). The consequence of such a situation is twofold. First, the primary loses its envelope (or most of it) and forms a BH through an SN with an NK (which may be a source of moderate eccentricities), not a direct collapse, which will be otherwise expected. Systems with high eccentricity are assumed to form through an NK during the formation of a BH, because prior to and during a long MT phase or a CE phase eccentric orbits are assumed to circularize rapidly, because of tidal interactions \citep[but for short interactions the circularization may not occur completely; e.g.][]{Eldridge0911}. Second, the secondary becomes rejuvenated and can be observed for longer than a star with similar mass which haven't experienced a mass gain (before going off the MS), which enhances the detection probability. nBHBs with massive companions represent a significant fraction of the \gaia\ sample ($\sim22\%$--$53\%$; Table~\ref{tab:massiveCOMP}) and typically have large separations ($a\gtrsim100\Rsun$; Figure~\ref{fig:dists_COMP}), as the companion's RL has to be large enough to accommodate it. Both high masses and large separations increase the detection probability by \gaia, but shorter lives ($\lesssim2\myr$) partially counteract this effect. Massive stars are expected to be relatively young ($t_{\rm age}\lesssim10\myr$), so they must originate from young Galactic components like the thin disk, which agrees with the typical mass of BHs ($7-8\Msun$, Figure~\ref{fig:dists_COMP}), as expected for a solar-metallicity environment \citep[e.g.][]{Belczynski1005}.

Although OB stars dominate, low-mass stars ($\lesssim1\Msun$) and WDs are also represented in the \gaia\ sample, especially in nBHBs originating from old Galactic components and containing heavy BHs ($\Mbh>20\Msun$). Such companions are dim for most of their life span and thus visible only in the vicinity of the Sun ($\lesssim3\kpc$), where there is only a small fraction of the Galactic stellar mass. Additionally, among companions in nBHB progenitors, low-mass stars and WD progenitors ($\Mzamsb\lesssim8\Msun$) constitute a smaller fraction of the sample than heavier companions as was already found in \citetalias{Wiktorowicz1911}. On the other hand, the lifetimes of the lighter stars are much longer than OB stars, which gives a higher probability of being observed currently. What is more, they can originate from old Galactic components like thick disks, halos, or bulges, which constitute nearly 1/3 of the Galactic stellar mass.

In spite of being relatively rare in the \gaia\ sample, nBHBs with low-mass companions constitute the bulk of the \lamost\ sample. These are mainly MS stars with masses $\lesssim4\Msun$. Such stars have longer life spans than massive OB stars in the \gaia\ sample and therefore could have been formed in older stellar populations. Some of them have evolved off the MS and became giants in the CHeB phase. Then, their luminosities are on average higher than in the MS phase ($M_{\rm bol}\sim-0.9$ to $-4.4$) and so can be observed from higher distances, including the Galactic bulge ($D_{\rm max} \approx 6.8$--$9.5\kpc$). What is more, surface temperatures of giant stars are significantly smaller in comparison to MS stars owing to their larger radii, which makes the \lamost\ RV uncertainty lower (Bai et al., in preparation). On the other hand, a giant star's lifetime is much shorter than that of its MS predecessor.

\begin{deluxetable}{ccc}
    \tablewidth{\columnwidth}
    \tablecaption{Massive Companions}

    \tablehead{ Model & $N_{\rm nBHB, Gaia}$ & $N_{\rm nBHB, LAMOST}$}
    \startdata        
        \std     & \sci{1.1}{2} ($69\%$) & \sci{4.3}{-3}\hfill ($3.2\%$) \\
        \ssa     & \sci{1.2}{2} ($34\%$) & \sci{5.3}{-3}\hfill ($0.04\%$) \\
        \nkr     & \sci{2.9}{1} ($37\%$) & \sci{1.9}{-4}\hfill ($0.1\%$) \\
        \nkbe    & \sci{2.6}{1} ($33\%$) & \sci{1.4}{-4}\hfill ($0.08\%$) \\
        \imfs    & \sci{2.7}{1} ($67\%$) & \sci{1.1}{-3}\hfill ($2.8\%$) \\
        \imff    & \sci{3.4}{2} ($73\%$) & \sci{1.3}{-2}\hfill ($4.4\%$) 
    \enddata
    \tablecomments{Number of nBHBs in \gaia/\lamost\ samples that harbor massive companions ($\Mcomp>30\Msun$). Numbers in parentheses represent the fractions of the total sample.}
    \label{tab:massiveCOMP}
\end{deluxetable}

Although being strongly biased toward low-mass companions, \lamost\ can also detect nBHBs with massive companions (up to $\sim60\Msun$). As pointed out above, such binaries have to be wide. Counterintuitively, in the \lamost\ sample nBHBs are even larger ($a\gtrsim1000\Rsun$) than in the \gaia\ sample. Although larger separations give typically smaller RV variations, if the eccentricities are large ($\gtrsim0.2$; Figure~\ref{fig:dists_COMP}), the RV variations are potentially higher than in similar circular systems because the amplitude of orbital velocity depends on eccentricity as $\Delta v_{\rm orb}\propto (2+e^2)/(1-e^2)$. We note that an eccentric binary spends most of the time in slow phase, which may limit potential detections. On the other hand, massive stars have higher temperatures; thus, \lamost's RV variation measurements are less precise (Bai et al., in preparation). In the case of \lamost, nBHBs harboring companions with masses above $30\Msun$ are expected in $\lesssim5\%$ of all the detections (Table~\ref{tab:massiveCOMP}).

In contrast to those observable by \gaia\ and \lamost, the majority of nBHBs in the Galaxy actually have WDs as companions. WD progenitors $\Mzamsb\lesssim8\Msun$ are not a majority on the ZAMS if the mass ratio is assumed to be flat \citepalias{Wiktorowicz1911}, however, WDs are very long-lived, whereas heavier stars either merge with BH primaries or quickly form a second compact object, which results in binary disruption or formation of a double compact object \citep[double compact objects are undetectable by \gaia\ and \lamost\ and thus are beyond the scope of this paper, although we note that a BH+NS merger may produce radiation in the optical band; e.g.,][]{Metzger1912}. For example, if $\Mzams\gtrsim1.3\Msun$, the WD phase is longer than the MS phase for a star born $\sim10\gyr$ ago. Nonetheless, WDs compose only a small fraction of the observable sample because their low luminosities allow for detection only in the close vicinity of the Sun \citep[$d\lesssim25$--$1000\pc$;][]{Barstow1407}, i.e. $\lesssim1\%$ of the total stellar mass of the Galaxy (see Figure~\ref{fig:profiles}). Particularly, \citet{Torres0507} and \citet{Jimenez-Esteban1811} estimated that the completeness for WD detection by \gaia\ can be obtained only up to $100\pc$. These estimations agree with the rarity of WDs in nBHBs in the \gaia\ and \lamost\ samples estimated on the basis of our results.

\subsection{Separations and eccentricities}

The top right and bottom right panels of Figure~\ref{fig:dists} present distributions of separation and eccentricity, respectively. The \gaia\ sample is clearly skewed toward higher separations, but the \lamost\ sample, although having on average low separations ($\lesssim100\Rsun$), has a significant fraction of wide systems that are typically eccentric ($e\gtrsim0.2$). We note that higher separations not only are rare on the ZAMS but in realistic situations are also prone to disruptions or interactions with field stars \citep[e.g.][]{Klencki1708}.

In the case of \gaia, separations are limited to $\sim3000\Rsun$ owing to the imposed limit on the orbital period ($P_{\rm orb}\lesssim5\yr$) resulting from the mission life span. Wider orbits allow for easier astrometric detection of a companion's motion from larger distances. For example, with separation equal to $\sim2000\Rsun$, even the dimmest stars, for which \gaia's astrometric precision is the worst, are above the astrometric detection limit up to a distance of $15\kpc$. However, only the most luminous stars (O type) can be photometrically observed by \gaia\ from such distances. The preference for circular orbits, arising from tidal circularization occurring typically prior to the BH formation, ensures that the majority of projected orbits, i.e. projections of the orbit on the plane whose normal is parallel to the line of sight, of even highly inclined binaries have separations similarly as large as the original orbits. We note that the semi-major axis of the projected orbit is, in general, not equal to the projection of the semi-major axis of the original orbit. If the orbit is significantly eccentric, the projected orbit may have separation as small as the semi-minor axis of the original orbit. It is, therefore, possible that the separation of the projected orbit is below \gaia's detection limit, although the original orbit is large enough to be detectable. Such situations are included in our procedure (see Section~\ref{sec:nBHB_Gaia} for further details).

In contrast to \gaia, \lamost\ will continue to gather observations for longer-period binaries. Nonetheless, some reasonable limit should still be applied in order to obtain realistic estimates. For the purpose of this study, we assumed $P_{\rm orb}=10\yr$ as an upper limit for \lamost. Potentially, even longer-period nBHBs may be observable when the orbit is nearly covered with observations, or the surveys may be extended beyond the assumed $\approx10\yr$. In such a case, the predicted number of nBHBs detectable by \lamost\ would increase, but due to the lack of good constraints, we decided to stick to the value of $10\yr$. In the case of \lamost, eccentric binaries may be even easier to detect than circular ones owing to a higher difference in orbital velocity between pericenter and apocenter, but the necessary lack of interactions that may circularize the orbit, or the need for strong NK for a BH, makes their fraction small among detectable nBHBs. Eccentric binaries must also be wide ($\gtrsim1000\Rsun$); otherwise, the BH progenitor may fill its RL during the periapsis passage and the orbit will become circularized by tidal forces or CE evolution.

\subsection{Gravitational wave sources}

If the BH companion is massive enough to form an NS or a BH, an nBHB may become a double compact object (DCO). The lower mass limit is affected by the binary evolution (MT phases). Even if the companion has a suitable mass, an NK may prevent the formation of a DCO by disrupting the system. Here we present an analysis of DCO progenitors present in both the \gaia\ and \lamost\ samples. For a recent analysis of a broad population of DCOs in the Galaxy (including WDs) see, e.g., \citet{Breivik2007}.

\begin{deluxetable}{ccc}
    \tablewidth{\columnwidth}
    \tablecaption{DCO Progenitors among nBHBs}

    \tablehead{ Model & $N_{\rm nBHB, Gaia}$ & $N_{\rm nBHB, LAMOST}$}
    \startdata       
        \multicolumn{3}{c}{All}\\
        \std     & 8.3\hfill ($5.1\%$) & \sci{1.2}{-3}\hfill ($0.9\%$) \\
        \ssa     & 8.5\hfill ($2.5\%$) & \sci{2.1}{-3}\hfill ($0.03\%$) \\
        \nkr     & 2.5\hfill ($3.2\%$) & \sci{2.0}{-3}\hfill ($1.3\%$) \\
        \nkbe    & 1.6\hfill ($2.0\%$) & \sci{8.5}{-5}\hfill ($0.05\%$) \\
        \imfs    & 2.0\hfill ($4.8\%$) & \sci{2.9}{-4}\hfill ($0.8\%$) \\
        \imff    & 23\hfill ($4.9\%$) & \sci{3.3}{-3}\hfill\hfill ($1.2\%$) \\
        
        \multicolumn{3}{c}{Merging}\\
        \std     & $5.3$ ($3.3\%$) & \sci{7.4}{-4}\hfill ($0.6\%$)\\
        \ssa     & $5.9$ ($1.7\%$) & \sci{9.4}{-4}\hfill ($0.01\%$)\\
        \nkr     & $0.5$ ($0.7\%$) & \sci{5.5}{-5}\hfill ($0.04\%$)\\
        \nkbe    & $0.4$ ($0.6\%$) & \sci{3.5}{-5}\hfill ($0.02\%$)\\
        \imfs    & $1.3$ ($3.1\%$) & \sci{2.0}{-4}\hfill ($0.5\%$)\\
        \imff    & $16$\hfill ($3.5\%$) & \sci{2.1}{-3}\hfill ($0.7\%$)\\
    \enddata
    \tablecomments{Number of nBHBs in \gaia/\lamost\ samples that are progenitors of double compact objects. Numbers in parentheses represent the fractions of the total sample. Both the total population of double compact object progenitors ("all") and those that have time to merger lower than $10\gyr$ (``merging'') are shown.}
    \label{tab:dco}
\end{deluxetable}

Less than $\sim5.1\%$ of nBHBs detected by \gaia, or less than $\sim1.7\%$ by \lamost, are going to form DCOs (Table~\ref{tab:dco}). The rest will either merger or form a BH-WD system. We note that BH-WD systems are potential gravitational wave sources \citep[e.g.][]{Kremer1908}, but the time to merger is larger than the Hubble time unless the separation is very small ($\lesssim10\Rsun$). \citet{Nelemans0109} found through PS study that only three BH-WD mergers per year will be detectable above the noise for LISA in the Galactic disk. Recently, \citet{Breivik2007} argued that none of such systems will be found by this instrument during its $4\yr$ of operation either in the Galactic disk or in the bulge, adopting a signal-to-noise ratio limit of $\mathrm{S/N}>7$.

The small fraction of DCO progenitors is a direct consequence of the small fraction of massive  companions ($\gtrsim8\Msun$, thus DCO progenitors) in nBHBs which are expected to be observed. In the case of \gaia, most of the companion stars are massive enough to form an NS or a BH. However, there are two effects that limit the DCO formation efficiency for these nBHBs. First, most of the companions in these nBHBs are on their MS, which means that their main expansion phase is still before them. Massive stars may fill their RL during the HG and start an MT phase that, due to a typically high mass ratio ($q>3$), is unstable and leads to a CE phase. A CE phase with HG donors is expected to lead to a merger, because of the lack of a clear core--envelope boundary \citep{Ivanova0402}, what ends binary evolution. Even if the secondary fills the RL in a later evolutionary phase when its core is well developed, the CE may result in a merger if there is not enough orbital energy to eject its massive envelope. The second factor that limits the number of DCO progenitors for the \gaia\ sample is the NK. During formation of the second compact object, the kick may disrupt the system, leaving two single stars.

\begin{figure}
    \centering
    \includegraphics[width=\columnwidth]{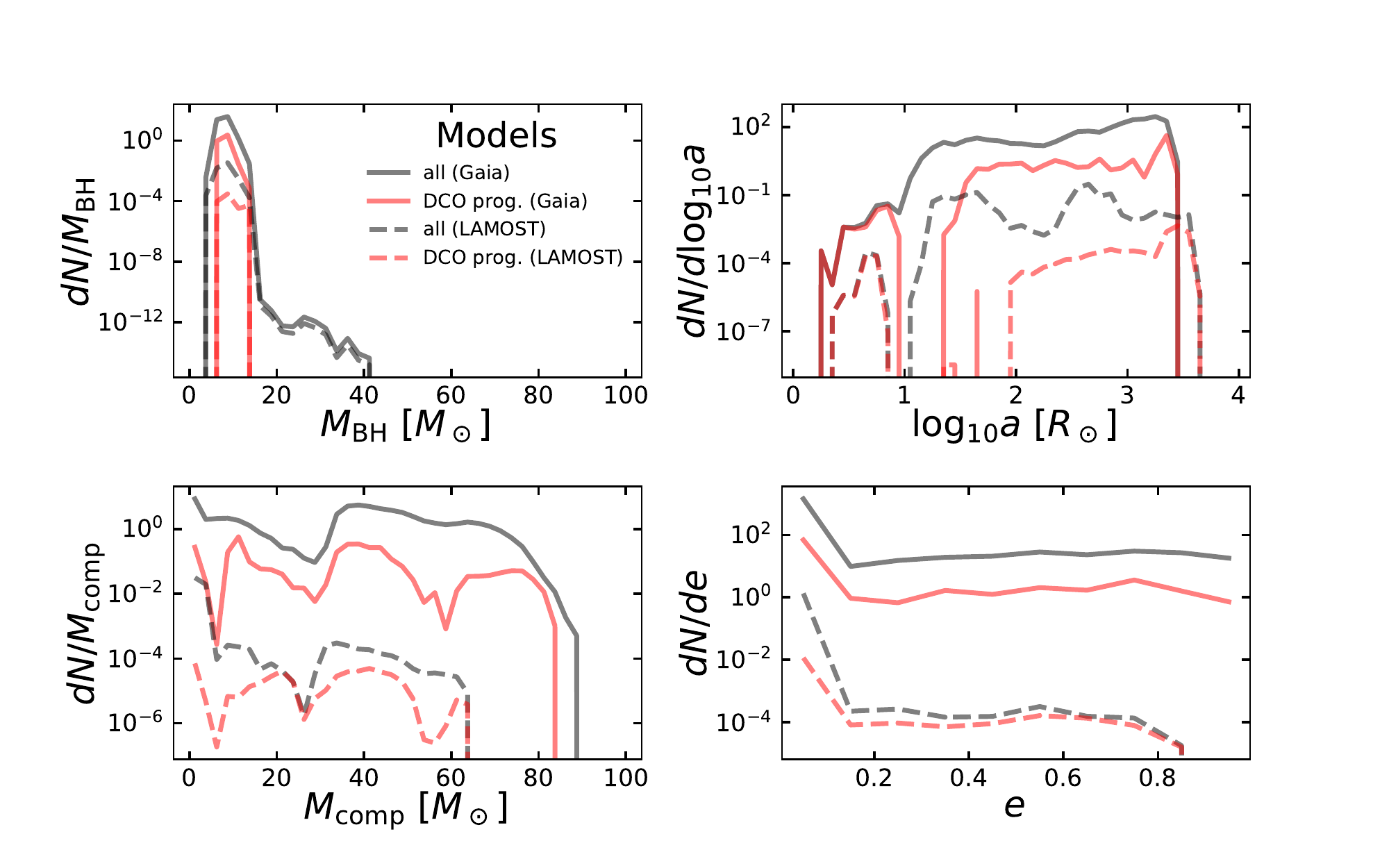}
    \caption{The same as Figure~\ref{fig:dists}, but for the \std\ model only and with separated distributions for DCO progenitors.}
    \label{fig:dists_dco}
\end{figure}

As far as the separation distribution is concerned, the DCO progenitors can be divided into two distinct groups. Those with small separations ($\lesssim10\Rsun$) contain helium stars that have lost their hydrogen envelopes as a result of interactions (typically CE) with BHs. These systems are close enough to survive even a strong NK during the formation of an NS where no fallback is expected. However, when the separation is higher ($\gtrsim10\Rsun$), the strong NK can easily disrupt the system. DCO progenitors with higher separations are typically MS stars that have not interacted with their primaries after the BH formation. Such separations are large enough to contain a massive ($\gtrsim8\Msun$) compact object progenitor. The fate of these systems is twofold. Either they will interact with their primaries as a result of nuclear expansion and lose their hydrogen envelope, effectively joining the systems with low separations and following similar evolution to that described earlier in this paragraph, or their RLs are large enough that the star may evolve without filling them and form a BH through direct collapse or with a small and/or well-directed NK avoiding disruption. The low-separation group typically contains progenitors of BH+NS systems, whereas the more numerous high-separation group contains progenitors of both BH+NS and BH+BH systems.

The formation routes of DCO progenitors are in general similar to the typical routes leading to the formation of nBHBs observable by \gaia\ and \lamost\ (Table~\ref{tab:evroutes}); thus, many DCOs existing currently in our Galaxy \citep[e.g.][]{Belczynski1006} evolved through the nBHB phase. The parameter distributions are similar in shape to those obtained for general population (Figure~\ref{fig:dists_dco}) except for the distribution of separations, as explained above. The lack of DCO progenitors with massive ($>20\Msun$) BHs results from the rarity of nBHBs with massive BHs in the total populations (Figure~\ref{fig:dists}). 

\begin{figure}
    \centering
    \includegraphics[width=\columnwidth]{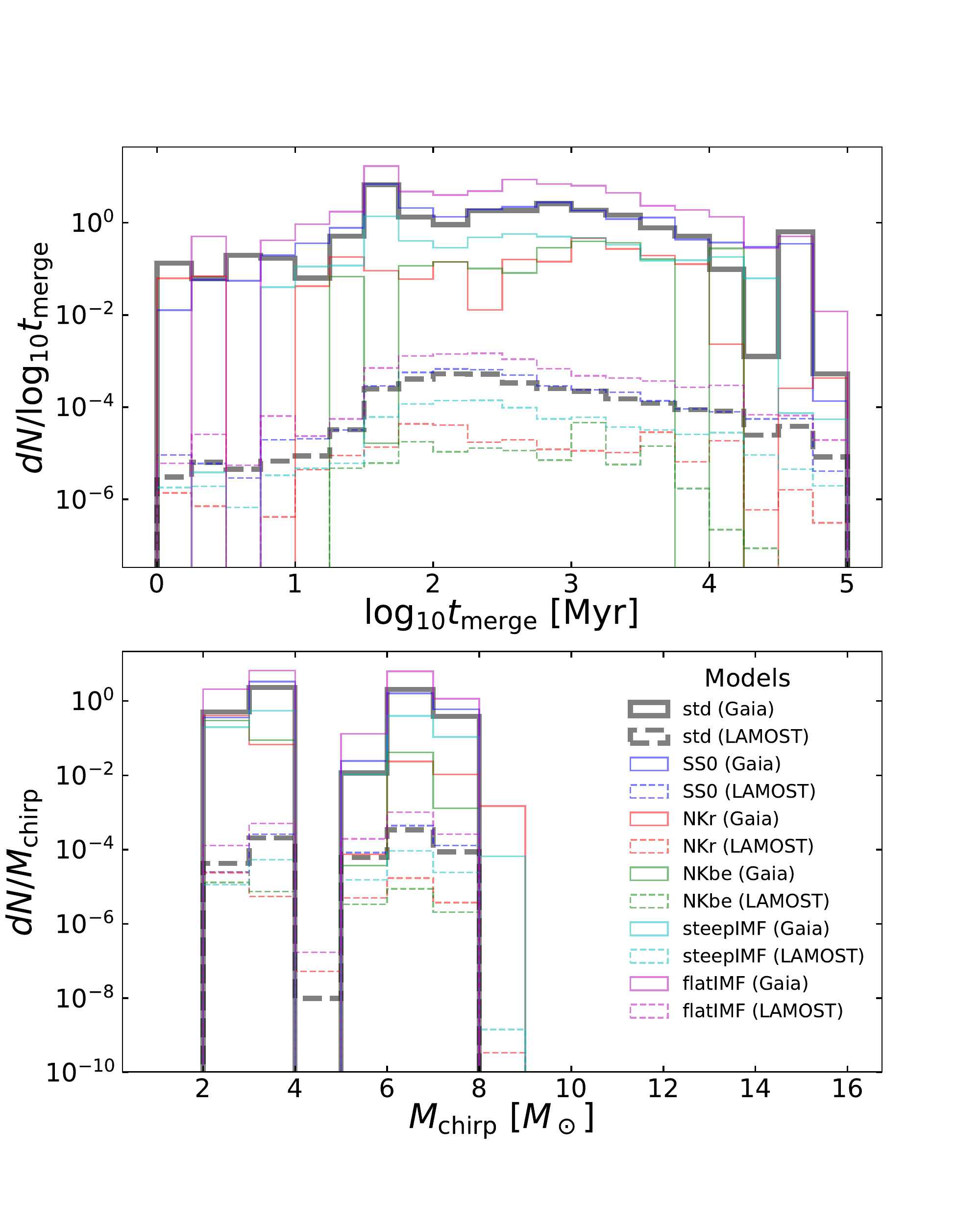}
    \caption{Distributions of time to merger ($t_{\rm merge}$) and chirp mass ($M_{\rm chirp}$) of DCOs which originate from nBHBs observed by \gaia\ and \lamost. The $t_{\rm merge}$ distribution was truncated at $10^5\myr$, which is a much longer time than the age of any Galactic component. Chirp masses are provided only for merging objects. The left peak at $\sim3\Msun$ corresponds to BH+NS systems, whereas that at $\sim6\Msun$ corresponds to BH+BH systems.}
    \label{fig:dists_dco2}
\end{figure}

The most interesting group among DCO progenitors are those that are going to merge, typically defined as DCOs with time to merger ($t_{\rm merge}$) lower than about $10\gyr$. Only a small fraction ($\lesssim5\%$) of nBHBs observable by \gaia\ or \lamost\ are expected to form DCOs, and only part of them will have $t_{\rm merge}<10\gyr$ ($\lesssim70\%$; see Table~\ref{tab:dco} and Figure~\ref{fig:dists_dco2}). The minimal $t_{\rm merge}$ in our results is $20\kyr$, so none of the DCOs that are expected to form from nBHBs that are expected to be observed by \gaia\ or \lamost\ can merge within this time. What is more, massive stars in the \gaia\ sample are typically still on their MS, whereas evolutionary advanced companions, frequent in the \lamost\ sample, are generally low-mass, and thus long-lived, stars. Therefore, the time to DCO formation is generally even longer than $t_{\rm merge}$. However, there might be some nBHBs that are evolutionary advanced (e.g. nBHBs with helium star companions in the \gaia\ sample) that can form DCOs in the near ($\lesssim500\kyr$) future. Such DCOs will emit gravitational waves in the frequency range of $\sim 10^{-4}$ to $10^{-3}$\,Hz with lifetimes of a million years and can be detected by LISA \citep{Babak1705}, Taiji \citep{2019arXiv190907104R}, or Tianqin \citep{2019PhRvD.100d3003W}. 

\section{Discussion}\label{sec:discussion}

\subsection{Comparison with previous studies}

\citet{Breivik1711} was the first to apply detailed PS modeling to study the detactability of nBHBs by \gaia. They used {\tt COSMIC}, an updated {\tt BSE} code \citep{Hurley0202}, assuming a primary-mass-dependent binary fraction \citep{vanHaaften1304}. They included only the thin and thick disks, the former with a total stellar mass of  $\sci{2.15}{10}\Msun$ and constant star formation throughout the past $10\gyr$, and the latter with $\sci{2.6}{9}\Msun$ and burst-like star formation that occurred $10\gyr$ ago \citep{Robin0310}. The distribution of stars was adopted after \citet{Yu1504}. Using these assumptions, \citet{Breivik1711} obtained a prediction of $(3.8$--$12)\times10^3$ detections during \gaia's $5\yr$ life span. Our model \ssa\ is the one that is the most similar to their assumptions. We obtain an order of magnitude fewer nBHBs ($\sim340$) in our calculations than their lower limit. The main reason why they obtain much more predicted nBHBs is the lack of any treatment of interstellar extinction included in their analysis, which significantly decreases the volume from which low-mass stars (the majority in their sample) can be observed. Especially, extinction makes it impossible to detect low-mass stars from the vicinity of the Galactic bulge. Other significant differences between their and our analysis, which does not necessarily increase the predicted number of nBHBs, include the lower total stellar mass ($\sim\sci{2.4}{10}\Msun$), a continuous relation between the binary fraction and primary mass ($0.5+0.25\log m$), and a lower limit for initial orbital periods of $0.5$ days.   

\citet{Shao1909} also recently estimated the number of nBHBs observable by \gaia. They performed a study of BHBs with normal-star companions (which they define as a star on the MS or at the (super)giant stage). Similarly to \citet{Breivik1711}, they used the binary PS code {\tt BSE} to simulate the Galactic population of BHBs. In their results, the pre-SN primaries are always helium stars that are assumed to have a probability of $0.6$ to become a BH if their mass is larger than $5\Msun$. The assumed total stellar mass in the Galaxy is $\sci{3}{10}\Msun$, which is roughly half of the stellar mass assumed in our Galaxy model, and with a constant SFH. Although \citet{Shao1909} included only binaries with initial periods $P_{\rm orb}<\sci{10}{3}$ days, in contrast to \citet{Breivik1711}, they included interstellar extinction to make the estimate more realistic. They predicted that the population of nBHBs with MS or giant companions observable by \gaia\ is $\sim262$--$974$ depending on the model. This number is comparable to our model \ssa, which is most similar to theirs as far as initial populations are considered. We note that, although their total stellar mass is only about half of what we use in our analysis, they allowed for stable MT for mass ratios as high as $\sim6$, assumed that all stars are in binaries, and in one model (B) allowed for a formation of BHs from stars with initial masses as low as $\sim15\Msun$. All these assumptions increase the number of expected nBHBs.

For \lamost\ we predict only $\lesssim14$ detectable nBHBs, which is so low partially because its field of view covers just $\sim6\%$ of the Galaxy. Our study is the first for this instrument that takes into account binary evolution. \citet{Yi1912} performed a simplified analytical calculation modeled after \citet{Mashian1709}, and obtain a prediction of $\sim50$--$400$ nBHBs depending on the survey strategy. This number is significantly higher than our most promising estimations (the \ssa\ model, with $14$ expected detections). The difference stems from the fact that their modeling lacks any treatment of interactions between binary components (like RLOF or CE), which, as we show in this study, are prevalent among progenitors of nBHBs in the predicted \lamost\ sample. On the other hand, they limited themselves to MS companions only and predicted that they are mostly low-mass stars on close orbits ($P_{\rm orb}\approx0.2$--$2$ days). The masses in general agree with our prediction, but our periods are typically higher. We note that their choice of uniform distribution of inclinations is biased toward face-on orbits compared to random orbital orientations, which increases their expected number of nBHBs detectable by \lamost. 

\subsection{nBHB candidates}

\begin{figure*}
    \centering
    \includegraphics[scale=0.43]{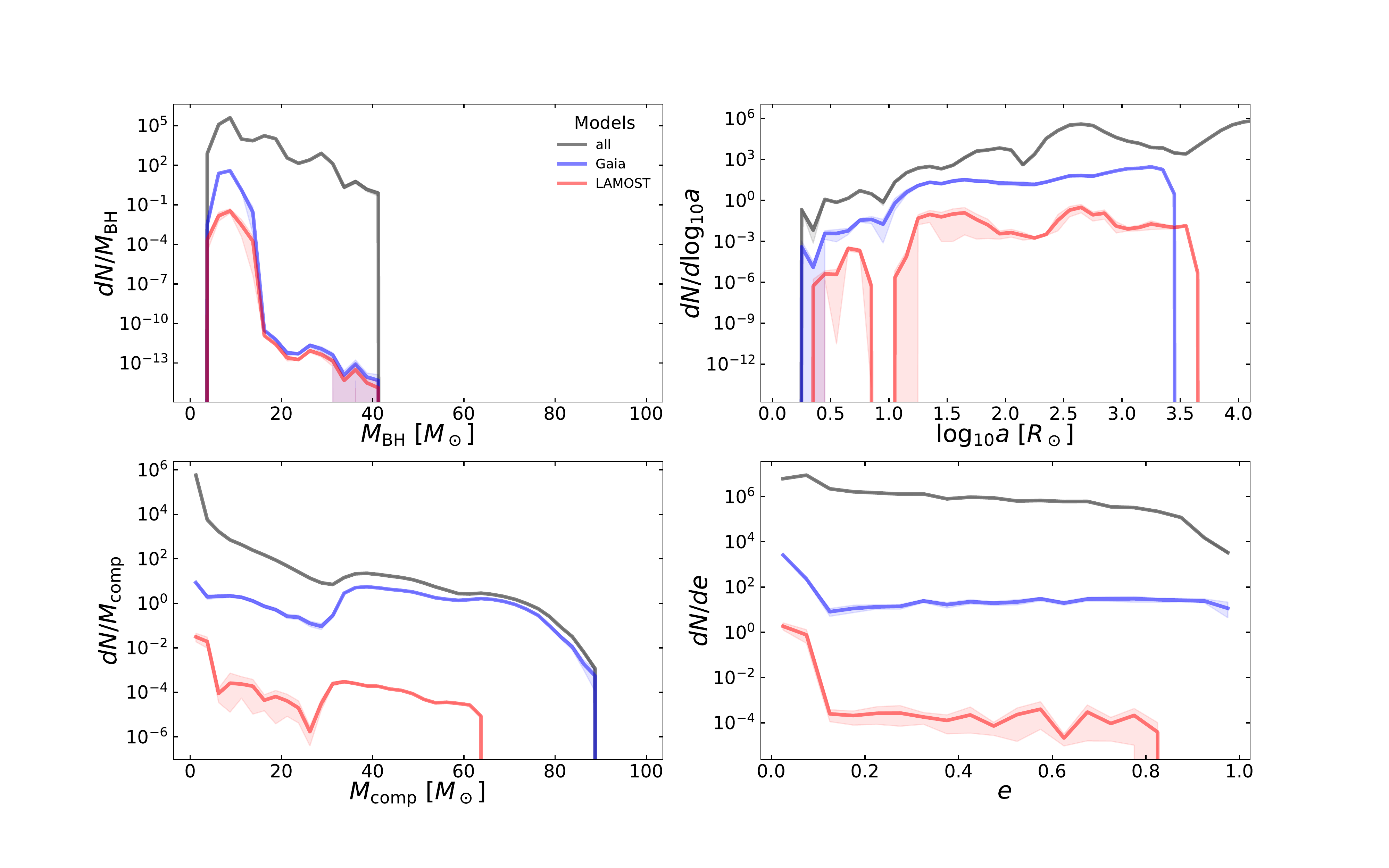}
    \caption{Similar to Figure~\ref{fig:dists}, but comparing the entire population of nBHBs (all) with the \gaia\ and \lamost\ samples. Only the \std\ model is shown for visibility reasons.}
    \label{fig:dists_all}
\end{figure*}

Although \gaia\ and \lamost\ are excellent instruments to search for nBHBs, the majority of the systems ($>99\%$) remain undetectable for them (Figure~\ref{fig:dists_all}). Up to date, no nBHBs have been detected through astrometry. In fact, only \gaia, with its unprecedented astrometric measurement accuracy, is able to provide a reasonable detection possibility, so we need to wait until the binary astrometric motion is going to be included in the forthcoming \gaia\ Data Releases or when the raw \gaia\ observations are available for in-depth study.

Despite many spectroscopic surveys \citep[e.g.][]{Trimble6906,Casares1401,Gu1902,Makarov1904,Zheng1911}, only a few strong candidates were identified. We note that all these candidates for nBHBs were specifically chosen for detailed analysis because of their luminous companions (RG with $m_{\rm V}=12.9$ or B-type stars with $m_{\rm V}=11$ and $11.5$) and high RV variations ($44\kms\leq K\leq53\kms$). \lamost\ is capable of detecting nBHBs with much smaller RV variations ($\lesssim10\kms$) and potentially fainter companions ($m_{\rm lim}\geq16$).

\citet{Khokhlov1804} analyzed a high RV variation  ($K\approx52\kms$) binary AS 386 with a B-type star and an invisible companion. The mass function provided a minimal mass for the latter as $1.9\Msun$. The position of the B-type star on HRD suggests a mass of $7\pm1\Msun$. This estimate, compared with the mass function, gives a minimal mass of the invisible companion of $7.3\Msun$ which makes it a plausible BH candidate. Although the evolution of the B star was probably affected by binary interactions (actually the Doppler tomography showed a presence of dust scattered around the B star, which supports this supposition), as it is typical for sources in the \lamost\ sample, so the actual mass may be different. Nonetheless, the presence of a BH in this system cannot be rejected.

Another nBHB candidate was detected by \citet{Thompson1911}, who analyzed a 2MASS J05215658+435922 RG star with strong RV variations ($K\approx44\kms$) and obtained a mass function of $0.77\Msun$. Although the orbit is relatively wide ($P_{\rm orb}=83$ days), the authors assumed (motivated by low eccentricity $e\approx0.005$ and similarity of orbital and photometric periods) that the giant's rotation is synchronized with its orbital motion and were able to assess its radius and imply a mass of the invisible companion of $\sim3.3\Msun$.

Recently, \citet{Liu1911} claimed the discovery of a BH accompanied by a $\sim8.2\Msun$ B-type MS star. Interpreting apparent motion of an H$\alpha$ emission line as associated with the BH further leads to a BH mass of $\sim70\Msun$. \citet{Belczynski2002} showed that such massive BHs may be produced through isolated evolution avoiding pair-instability SNe, when the stellar evolution is fine-tuned and for extreme mass-loss assumptions. However, in this study we adopted more conservative evolutionary physics, which could not reproduce such an object. Nonetheless, parameters of the visible companion were revised by many further studies \citep[e.g.][]{Eldridge1912,Simon-Diaz2002,El-Badry2003}, and especially \citet{Irrgang2001} suggested that the companion may be actually a low-mass ($1.1\pm0.5\Msun$) stripped star, which is mimicking a B-type MS star. Together with the updated estimation of the binary mass ratio \citep[$\sim 5.1\pm 0.8$;][]{2020arXiv200512595L}, the BH mass is lowered to $\sim4$--$36\Msun$. \citet{Yungelson2004} performed a PS study based on Irrgang et al. finding that such systems can exist in the Galaxy in significant numbers. However, they have not included observability for any instrument in their calculations. We note also that recently \citet{Shenar2004} showed that by disentangling the spectra it can be claimed that the binary actually harbors no degenerate components.
This discord motivated us to briefly analyze what is the probability of detecting nBHBs with B-type companions in the \gaia\ and \lamost\ samples aiming at reproducing the LB-1, but retaining a significant parameter freedom. Specifically, we analyzed the presence of nBHBs with hot companions ($10$ kK $<T_{\rm eff}< 14$ kK) and orbital periods of $80\pm10$ days in our \gaia\ and \lamost\ samples. We note that hot stars have significant RV errors in \lamost\ observations, which can hinder detection. Such binaries must have gone through interactions like CE or MT, because the orbit is too small ($a\approx300\Rsun$) to accommodate the BH progenitor, and if the companion is indeed a low-mass star, it can obtain such high temperatures only if their envelopes were stripped off through RLOF. The interaction phase is supported by the detection of surface molecule anomalies that may originate from mass stripping revealing processed material. The expected number of such binaries is $0.18$ in the Gaia sample and $2\%$ of the \lamost\ sample for the STD model. Such systems have B-type MS stars, which went through a CE phase, reducing the initial separation of $\sim5000\Rsun$ to just $\sim20\Rsun$. When the system is $\sim100\myr$ old and consists of a $\sim7\Msun$ BH and a young $\sim4$--$5\Msun$ B-type MS star, it becomes observable by both \gaia\ and \lamost. These numbers are much smaller than obtained by \citet{Safarzadeh1912}, who calculated $\sim60,000$ LB-1-like systems on the basis of the \citet{Liu1911} observation, and $\lesssim4000$ on the basis of their analysis. The discrepancy originates from many simplifying assumptions adopted by \citet{Safarzadeh1912}, like equal lifetimes of $>8\Msun$ stars, constant SFH, no super-Eddington MT rates, no extinction, etc., and considering only one evolutionary path leading through a ULX phase (we show that nBHBs can form without an XRB (let alone ULX) phase; see Table~\ref{tab:evroutes}). 

To close this subsection, we note here that the predicted number of nBHBs that can be detected by surveys such as \gaia\ or \lamost\ are model dependent. Different models may result in very different numbers, which again suggests that searching for nBHBs using astrometry or RV surveys may put strong constraints on the evolutionary models, their initial conditions, and the structure of the Galaxy.

\subsection{Milky Way Model Influence}

\begin{figure}
    \centering
    \includegraphics[width=\columnwidth]{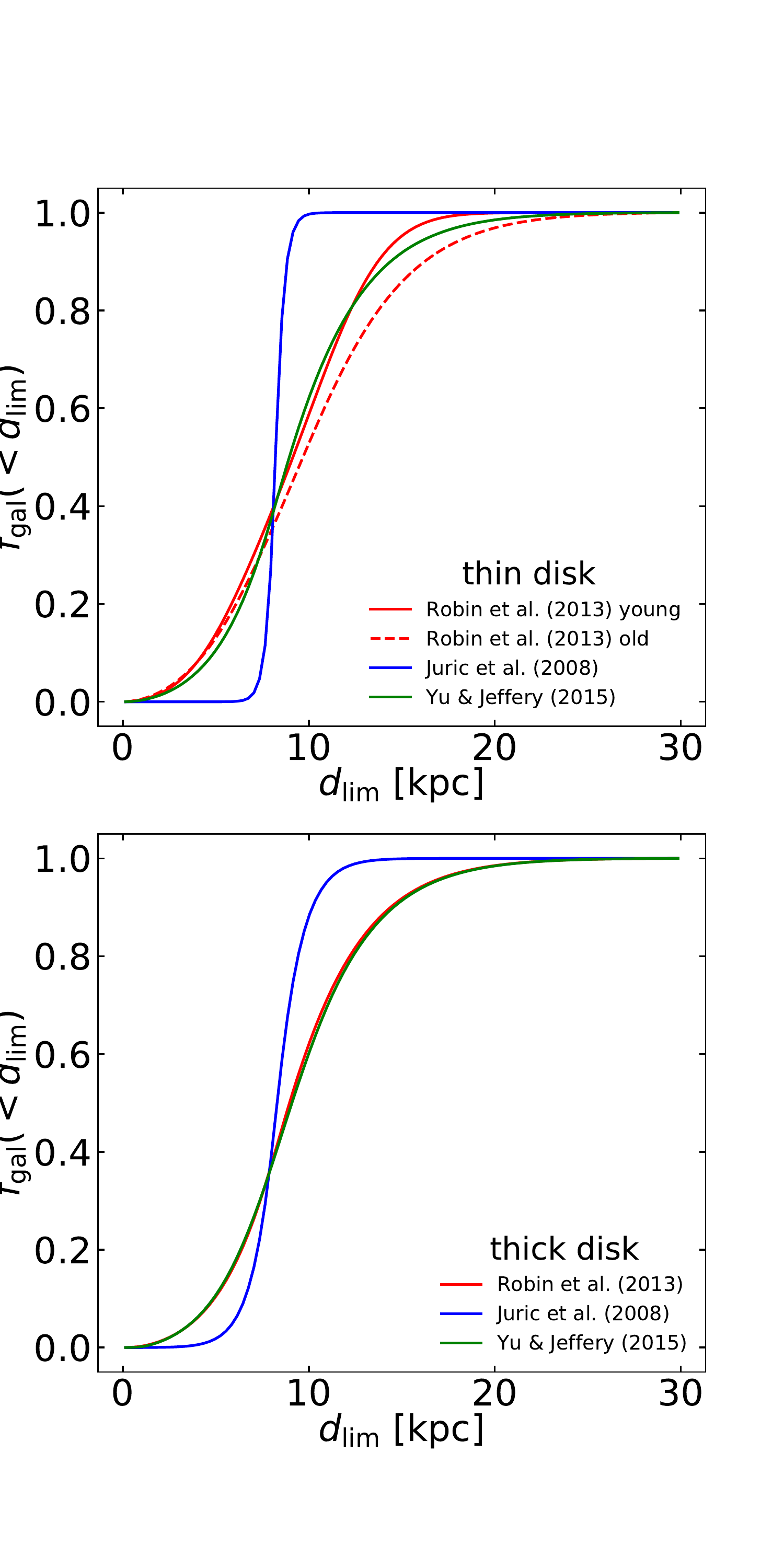}
    \caption{Same as Figure~\ref{fig:profiles}, but different models of stellar distributions are compared and only thin (top) and thick (bottom) disks are presented.}
    \label{fig:profiles_comparison}
\end{figure}

The spatial distribution of stars in the Galaxy is still very uncertain. In general, it is much harder to estimate global parameters like color or mass for the MW than for other galaxies, because we cannot make an unbiased observation ``from inside.'' Here we compare the model utilized in this study with other available models in order to estimate the influence the MW model has on our results. Figure~\ref{fig:profiles_comparison} presents a comparison between the models for the thin and thick disk, as most of the other studies concentrated on these two components only.

\citet{Mashian1709} and \citet{Yi1912} utilized the double exponential model of \citet{Juric0802} in which $dN/dV\propto \exp(-R/L-Z/H)$, where $R$, $Z$ are cylindrical galactocentric coordinates and $L$, $H$ are constants. Figure~\ref{fig:profiles_comparison} shows that this model gives $f_{\rm gal}(d_{\rm lim})$ very different from other models and, especially, shows a much higher concentration of stars near the Galactic center. 

\citet{Breivik1711} employed the model of \citet{Yu1504}. This model assumes an exponential relation for the distance from the Galactic center ($R$) but uses hyperbolic secant for the distance from the Galactic plane ($z$). As a result, the $f_{\rm gal}$ relation is very similar to the one that we employed in this work, and we expect no significant differences in our results if the model of \citet{Yu1504} was applied for the thin and thick disks. 

Another important factor that is not well constrained for the Galaxy is its SFH. Here we assumed that the star formation rate was constant during the start formation periods for all Galactic components, as shown in Table~\ref{tab:MWcomponents}. However, such an approach may significantly overestimate the recent star formation in the Galaxy. According to our model, the recent star formation in the thin disk is $4.7\msy$. However, \citet{Robitaille1002} estimated a much lower value of $0.68-1.45\msy$. Also, \citet{Mutch1108} suggested that the MW is in the state of ceasing star formation and becoming a red spiral galaxy, which suggests a lower recent star formation than average. The lower value of recent star formation will decrease the fraction of massive stars that are present in current stellar populations owing to their short lifetimes. As a consequence, our prediction for the \gaia\ sample, which consists mostly of nBHBs with massive companions, would decrease by a factor equal roughly to the recent star formation divided by the star formation assumed in this study ($\mathrm{SF}_{\rm recent}/4.7\msy$). If the recent star formation is indeed $\lesssim1\Msun$, it would mean that the expected number of detections should be lower by a factor of $\gtrsim4$--$5$ compared to the values estimated in this study (Table~\ref{tab:results}).

What is more, the metallicity and its dependence on location and Galactic age are not known precisely for the MW. The metallicity significantly influences the evolution of stars, especially the mass loss in stellar wind and radial expansion \citep[e.g.][see also \citeauthor{Vos2003}~\citeyear{Vos2003}]{Belczynski0706,Belczynski1006}, which plays a crucial role on the final BH mass in an nBHB. Better constraints can significantly alter theoretical predictions.

\begin{figure*}
    \centering
    \includegraphics[width=\textwidth]{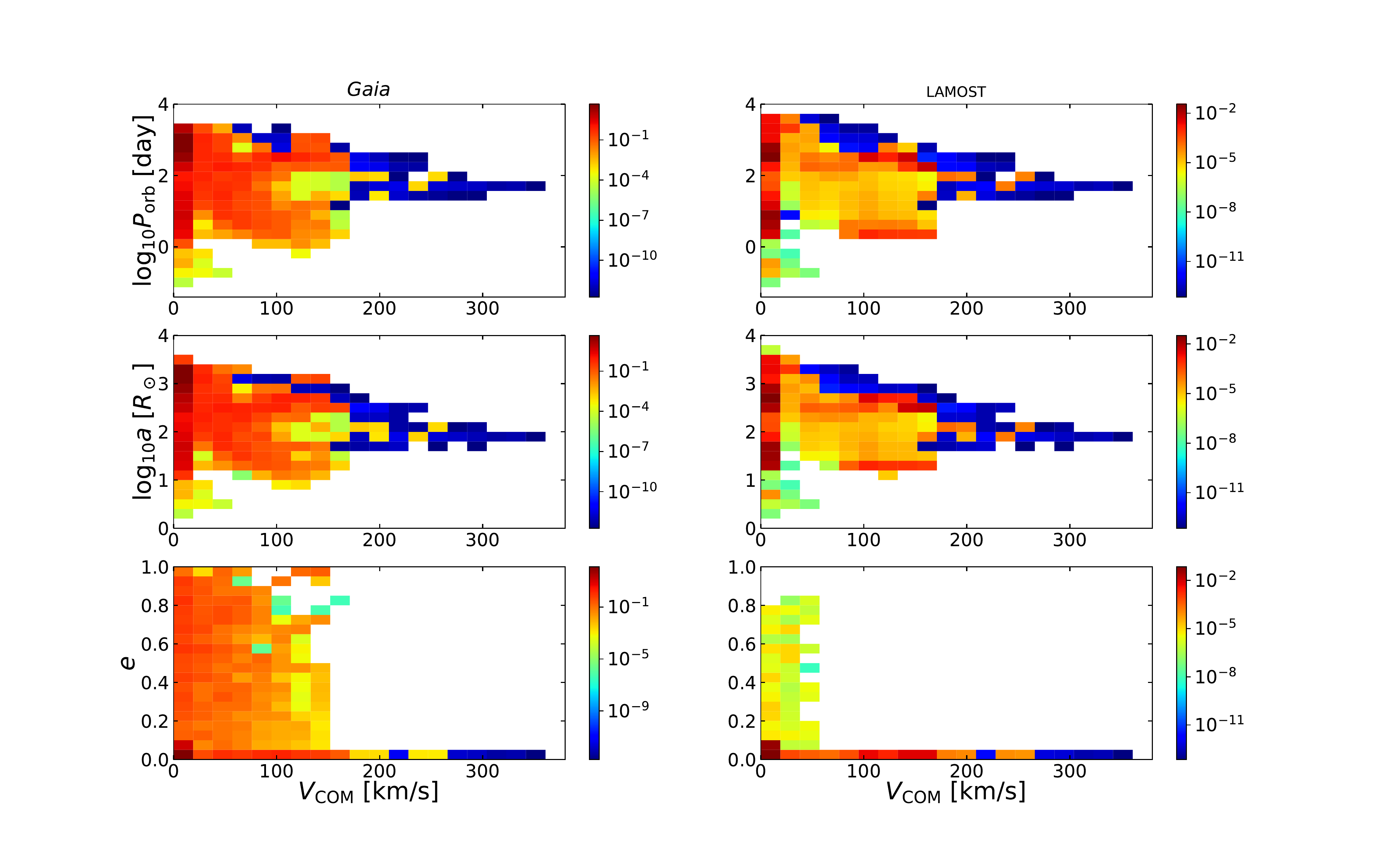}
    \caption{2D histograms of orbital period ($P_{\rm orb}$) vs. center-of-mass velocity ($V_{\rm COM}$; top panels), separation ($a$) vs. $V_{\rm COM}$ (middle panels), and eccentricity ($e$) vs. $V_{\rm COM}$ (bottom panels). Results are presented for the \gaia\ (left panels) and the \lamost\ (right panels) samples and the \std\ model.}
    \label{fig:Porb_Vcom}
\end{figure*}

In this paper we have assumed that the spatial distribution of nBHBs follows the distribution of stars. However, in realistic situations, nBHBs can have different distributions owing to NKs that can change not only binary parameters but also the center-of-mass velocity. Compact binaries ($a\lesssim10\Rsun$) can obtain velocities as high as $\sim300\kms$ while remaining bound. Therefore, the distribution of compact nBHBs can be much more extended (dynamically hotter) than the distribution of stars. These binaries are not easily detectable by \gaia\ owing to their smaller sizes, whereas \lamost\ has worse accuracy for hotter stars (the result of interaction). Nevertheless, compact nBHBs with large center-of-mass velocities are expected to be rather rare. Wider pairs with slow orbital motions, which are a majority of both the predicted \lamost\ and \gaia\ samples, are prone to disruption if the compact object formation is associated with a significant NK. Binaries that survive the process would acquire only small additional velocities, which means that they will in general follow the distribution of stars as assumed in this work. We note that wide nBHBs are likely to interact with other stars even in the Galactic field \citep{Binney08,Klencki1708}, which can lead to disruptions or increase in eccentricity that effectively leads to shortening of the orbital period.

In Figure~\ref{fig:Porb_Vcom} we present 2D histograms of center-of-mass velocities and orbital periods. The majority ($>80\%$) of systems have small ($<20\kms$) velocities, and only systems with periods less than $100$ days obtain significant ($\gtrsim250\kms$) velocities. The same relation is observed in the \gaia\ and \lamost\ samples. In addition, we show the distribution of eccentricities in relation to center-of-mass velocities to support the statement that high-velocity systems ($\gtrsim150\kms$), which are not wide ($<700\Rsun$), quickly circularize ($e\approx0$), whereas wide systems ($>1000\Rsun$) can be eccentric (up to $e\approx1$).

\subsection{Detection probability}

Although about $100$ nBHBs are predicted to be observed by \gaia\ during its $5\yr$ long mission, not all of these systems may be recognized as such, due to projection effects. For example, a projected eccentric binary may have a semi-major axis ($a_{\rm proj}$) smaller than the semi-major axis of the original orbit. It may not only make the astrometric motion of the nBHB unobservable by \gaia, as included in our study, but also hinder the posterior analysis, which is necessary in order to obtain the invisible object's mass. 

Provided that the orbital orientation and companion's mass can be inferred from the observations, the hidden object's mass can be calculated from the equation \citep[see also][]{Andrews1909}:
\begin{equation}
    \frac{M_{\rm prim}^3}{(M_{\rm comp}+M_{\rm prim})^2}=\frac{4\pi^2}{G P_{\rm orb}^2}a_{\rm comp}^3,
\end{equation}
\noindent where $G$ is the gravitational constant and $M_{\rm prim}$ is the mass of a hidden primary (a BH candidate in our case). \gaia\ will observe every object on average $\sim75$ times, which may provide enough data to constrain the orbital orientation. Otherwise, $a_{\rm proj}$ gives only a lower limit for $a_{\rm comp}$. Nonetheless, nBHBs observable by \gaia\ are expected to by nearly circular in most cases, because the tidal interactions during the MT phase proceeding the BH formation should circularize even very eccentric orbits, unless the NK was very strong and the orbit is wide. Therefore, as a first approximation we may assume $a_{\rm comp}\approx a_{\rm proj}$ for nBHB candidates. \citet{Andrews1909} calculated that, for a companion mass of $10\Msun$, nBHBs with a BH of mass $10\Msun$ should be detectable at least to a distance of $1\kpc$. We find that heavier companions are expected to form a bulk of  the \gaia\ sample and they are observable mostly from the vicinity of the Galactic bulge, which goes beyond their tested distance range. 

The other significant problem is the mass of the visible star. In many cases it can be estimated by fitting its spectra to those calculated for single stars, e.g. from PARSEC tracks or single-star PS. We note that not only are such fits typically not very precise, but also the evolution of a star in a binary system can be affected by interactions with the companion. Actually, for nBHB progenitors such interactions are expected to be typical (Table~\ref{tab:evroutes}). Fortunately, in most cases the MT occurs when the companion is still on its MS. For MS stars, the MT results in rejuvenation of the accretor, because it has enough time to adapt to the increase of its mass and continue its evolution similarly to a single star of accordingly higher mass.  

In general, or when the orbital orientation and companion mass have not been derived from the observations, we can estimate the absolute lower limit on the mass of the invisible companion as
\begin{equation}\label{eq:m_prim}
    M_{\rm prim}>M_{\rm prim, min}=\frac{4\pi^2}{G P_{\rm orb}^2}a_{\rm proj}^3.
\end{equation}
\noindent If $M_{\rm prim, min}$ is higher than $2.5$--$3.5\Msun$ (theoretical upper limits for the NS mass) and, simultaneously, the distance to the system is lower than the distance at which a ZAMS star of mass $M_{\rm prim, min}$ will have apparent magnitude equal to \gaia's limiting magnitude of $21$ mag (stars generally tend to increase their luminosity during their evolution), the system contains a plausible BH candidate.

We note that $P_{\rm orb}<5\yr$ is a limit imposed by the mission duration, not by any physical constraints. Even binaries with longer orbital periods can be detected without their orbit being fully covered by observations, and their parameters will be constrained \citep{Lucy1403,Docobo1805}. The ability to detect longer orbits may significantly increase the detection rate of nBHBs by \gaia.

The bolometric correction for red supergiants (RSGs; $T_{\rm eff}\lesssim4000$ K) can be significant \citep{Jordi1011}. However, RSGs constrain only a small fraction of the \gaia\ sample. First of all, the RSG phase is very short in comparison to the MS or WD phases, and solar-metallicity massive stars ($>40 \Msun$) never expand significantly enough to become SGs owing to significant mass loss. Second, RSG progenitors reach radii of hundreds to thousands of $\Rsun$ during their evolution. For high-metallicity stars this occurs by the time the star finishes CHeB and is able to reach high central temperatures necessary for large luminosities \citep[e.g.][]{Wiktorowicz1911,Klencki2004}. Additionally, RLOF can remove the outer layers of the stars, raising their effective temperatures. On the other hand, recent simulations \citep[e.g.][]{Laplace2005} show that the hydrogen envelope may not be fully removed, and the star's remaining envelope can still expand significantly. To sum up, if the bolometric correction is applied to our results this should not influence significantly the predicted \gaia\ sample, although we agree that the values we provide should be considered as an upper limit.

Interpreting the \lamost\ spectroscopic observations suffers from a similar problem to that of \gaia's astrometric measurements. The observed RV variations depend on the inclination, \begin{equation}
K=50.6 \kms\;\sin i\;(a_{\rm comp}/\Rsun)(P_{\rm orb}/{\rm day})^{-1}(1-e^2)^{-0.5}.
\end{equation}
\noindent So RV variations from any nBHB visible face-on (inclination $i\approx0$) will be undetectable, and a mass function calculated from the orbital period and observed RV variations gives only the lower limit for the invisible star's mass. Additional measurements that allow estimating the orbital orientation (in the case of \lamost, only inclination is needed) and the visible star mass are necessary to constrain the invisible object mass. 
If this is not possible, we are left only with a lower limit coming from the mass function, which allows us to indicate BH candidates using a similar argument as for \gaia\ (see Equation~\ref{eq:m_prim} and the related text), but with the limiting magnitude appropriate for \lamost\ ($m_{\rm lim}=16$ for the current survey).

In our analysis, we assumed that \lamost\ will be able to observe all binaries that are within detection limits (which is the \lamost\ sample). However, the number of fibers limits the number of sources that are observed each night, and the number of measurements for a single source is limited. For the current low-resolution survey, the majority of objects are being observed only three times during the same night (cadence of $40$--$120$ minutes), which limits the periods for which RV variations can be detected to $\lesssim1$ day. Such binaries constitute only a small fraction of the predicted observable \lamost\ sample ($a\lesssim10 R_\odot$; see Figure \ref{fig:dists}, top right panel). The situation can change for the upcoming medium-resolution survey, in which the majority of objects are going to be observed $\sim60$ times during a more than $5\yr$ long observational campaign. Nevertheless, detected RV variations, unless a RV curve can be fitted, can give only lower limits for the intrinsic values of $K=v_{\rm rot} \sin i$. Therefore, some observable binaries, which are included in the \lamost\ sample, may remain undetected owing to rare data coverage. Our evolutionary models, supported by statistical analysis, find that typical spectroscopic binaries in the \lamost\ sample have large separations (and thus large orbital periods, $P_{\rm orb}\gtrsim1$ yr; Figure \ref{fig:dists}). Therefore, regular observations, even if sparse, can prove more efficient for nBHB detection than one-night short-cadence ones. On the basis of our results we can deduce that, if we focus on finding nBHBs and can make only a few observations of one object, a recurrent observations every $\sim1\yr$ that can detect RV variations for typical nBHBs in the \lamost\ sample are a more promising way to detect these sources, which can be realized in forthcoming surveys. 

\section{Summary}\label{sec:summary}
   
Using the recently published database of model predictions for BHs in different stellar environments \citepalias{Wiktorowicz1911}, we have constructed a population of nBHBs in the Galaxy using realistic models of the MW's SFH, chemical evolution, and spatial distributions of stars in different Galactic components (thin and thick disks, halo, and bulge). We applied observational constraints to theoretical evolutionary models in order to calculate the detection rates and properties of nBHBs detectable by \gaia\ and \lamost. In such a way we can use observations to constrain the theory. To account for uncertainties in stellar and binary evolution, we analyzed a range of realistic models that provide a better insight into the importance of evolutionary processes on the predictions. To give a broader picture, we discussed how future missions can improve our detection rates of nBHBs.

Specifically, we estimated the total number of BHs in nBHBs present in the Galaxy today as \sci{1.5}{5}--\sci{3.2}{6} depending on the evolutionary model employed. However, only a small fraction ($\lesssim0.05\%$) of these nBHBs can be detected by spectroscopic or astrometric observations of companion stars. We predict that \gaia\ is able to detect $\sim41$--$340$ nBHBs within $5\yr$, whereas \lamost\ is expected to detect less than $14$ nBHBs in a span of $10\yr$. The significantly lower detection rate for \lamost\ comes mainly from the fact that its field of view encompasses only $\sim6\%$ of the Galaxy's stars.

If the possibility of detecting incomplete orbits is taken into account \citep[e.g.][]{Lucy1403}, the numbers may be significantly larger, as many nBHBs are predicted to have orbits wider than $5$--$10\yr$. On the other hand, the poorly constrained SFH, especially recently, significantly affects the estimates. For example, if the recent star formation is as low as $\sim1\Msun{\rm yr}^{-1}$ \citep[e.g.][]{Robitaille1002}, the estimated numbers of nBHB detections for \gaia\ will drop to $\sim10$--$70$. 

Although the typical predicted masses of BHs in nBHBs are similar for both instruments ($\sim7$--$8\Msun$), typical companion masses differ significantly. In the predicted \gaia\ sample, massive OB stars with masses above $30\Msun$ dominate, whereas for \lamost\ these are mostly low-mass ($\lesssim4\Msun$) MS stars. We note that the separations in nBHBs observable by \gaia\ are expected to be much larger ($\sim1000$--$2000\Rsun$) than those by \lamost\ ($\sim10$--$1000\Rsun$), mainly due to the different requirements for the detection.

These predictions depend on evolutionary models encapsulated in PS results and models of the Galaxy, both of which can introduce significant errors that are hard to estimate.
However, the method we have developed is flexible enough to include future advancements in stellar and binary evolution and the MW composition and history.

We show how important are the treatments of binary interactions, SFH, metallicity, and spatial distribution of stars for the predictions of observational properties of nBHBs. The majority of Galactic stars are located in the vicinity of the bulge ($D\approx8\kpc$), and therefore observable only if their luminosities are higher than $\sim400\Lsun$. These stars can be very young as a part of the thin disk and typically have $\sim$solar metallicity. Less luminous stars are potentially detectable only in the vicinity of the Sun, where the stellar number density is much smaller. Stars of low metallicity are typically old and can be observed mainly in thick disks and halos, where only a small fraction ($\sim10\%$) of Galactic stars are located. Consequently, inclusion of realistic Galactic properties can significantly change the estimates in comparison to simplified models, especially when old or massive stars are involved. In our work we used recent estimates for these properties and adopted a novel method to include them in population studies. Future detections of nBHBs by \gaia, \lamost, and other surveys may put strong constraints on binary star evolution and PS models.

\acknowledgements

We thank Gijs Nelemans for a helpful suggestion. We are thankful to thousands of volunteers who took part in the {\it Universe@Home} project\footnote{https://universeathome.pl/} and made the creation of the database used in this study possible. GW is partly supported by the President’s International Fellowship Initiative (PIFI) of the Chinese Academy of Sciences under grant no.2018PM0017 and by the Strategic Priority Research Program of the Chinese Academy of Science Multi-waveband Gravitational Wave Universe (Grant No. XDB23040000). This work is partly supported by the National Natural Science Foundation of China (Grant No. 11690024, 11873056, and 11991052) and the National Key Program for Science and Technology Research and Development (Grant No. 2016YFA0400704). SJ acknowledges funding by the European Union’s Horizon 2020 research and innovation programme from the European Research Council (ERC; Grant agreement No. 715063, PI: de Mink) and by the Netherlands Organisation for Scientific Research (NWO) as part of the Vidi research program BinWaves (project number 639.042.728, PI: de Mink). KB acknowledges support from the Polish National Science Center (NCN) grant Maestro (2018/30/A/ST9/00050).

\bibliographystyle{apj}
\bibliography{ms}

\end{document}